\documentclass[aps,reprint,preprintnumbers,superscriptaddress,floatfix,nofootinbib,longbibliography,balancelastpage]{revtex4-1}
\usepackage[utf8]{inputenc}
\usepackage{textgreek}

\usepackage{comment}
\usepackage{dcolumn}
\usepackage{xcolor}
\usepackage{youngtab}
\usepackage{ytableau}
\usepackage{tabularray}
\usepackage{booktabs}    % some options for tabulars\\
\usepackage{placeins}
\usepackage{soul}
\usepackage{bm}
\definecolor{redd}{rgb}{0.8, 0.1,0.2}
\definecolor{navy}{rgb}{0.05, 0.23,0.75}
\usepackage[colorlinks=true,citecolor=red,linkcolor=blue]{hyperref}
\usepackage{chngcntr}
\hypersetup{colorlinks,	linkcolor={green!47!black},	citecolor={green!60!black},	urlcolor={green!55!black}}
\usepackage{bbm}
\usepackage{amsfonts}
\usepackage{amsmath,amssymb}
\usepackage{mathrsfs}
\usepackage{epsfig}
\usepackage{graphicx}
\usepackage{wrapfig}                 
\usepackage{url}
\usepackage[nameinlink]{cleveref}
\usepackage{float}
\usepackage{color}
\usepackage{multirow}
\usepackage{lipsum}
\usepackage{enumitem}
\newcolumntype{L}{>{\centering\arraybackslash}m{1.5cm}}

\usepackage{enumitem}
\usepackage{chngcntr}
\usepackage{tikz}

\newcommand{\dSSB}{\textrm{d}\textrm{SB}}
\newcommand{\acrit}{\alpha^\textrm{crit}_{\textrm{SB}}}

\newcommand{\orcid}[1]{\href{https://orcid.org/#1}{	\raisebox{0.5\height}{\includegraphics[height=1.25ex,width=1.25ex]{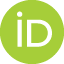}}}}

\newcommand{\commentmute}[1]{} %%mute

\graphicspath{{Figures/}}
\begin{document}
		
\preprint{RIKEN-iTHEMS-Report-26}

\title{Confinement~without~symmetry~breaking~in~chiral~gauge~theories}

\author{Hao-Lin Li\,\orcid{0000-0002-4310-1915}}
\affiliation{School of Physics, Sun Yat-Sen University, Guangzhou 510275, P. R. China}

\author{\' Alvaro Pastor-Guti\'errez\,\orcid{0000-0001-5152-3678}}
\affiliation{RIKEN Center for Interdisciplinary Theoretical and Mathematical Sciences (iTHEMS), RIKEN, Wako 351-0198, Japan}

\author{Shahram Vatani\,\orcid{0000-0002-1277-5829}}
\affiliation{ Centre for Cosmology, Particle Physics and Phenomenology (CP3), Université catholique de Louvain, Chemin du Cyclotron, 2 B-1348 Louvain-la-Neuve, Belgium}

\begin{abstract}
The infrared structure of gauge theories with chiral fermions remains largely unexplored. In this work we investigate the Bars--Yankielowicz class using the functional renormalisation group, building on recent developments in gauge--fermion systems that provide clear criteria for confinement and dynamical symmetry breaking. 
We show that two distinct phases arise: one exhibiting both confinement and symmetry breaking at small numbers of colours, and another characterised by confinement without symmetry breaking in the large-colour limit. The latter realises a novel regime, opening the possibility of exotic spectra and phenomena that can now be studied within a systematic framework.
\end{abstract}

\maketitle

\section*{Introduction}\label{sec:Introduction}
Understanding the infrared (IR) dynamics of four-dimensional gauge theories is one of the central challenges in modern quantum field theory (QFT). Significant progress has been achieved in physical quantum chromodynamics (QCD) and related theories with vector-like fermions, where a variety of exact and approximate tools are available. From the symmetry perspective, methods such as ’t Hooft anomaly matching conditions~\cite{tHooft:1979rat}, supersymmetric limits~\cite{Seiberg:1994rs,Seiberg:1994aj,Seiberg:1994bz,Seiberg:1994pq}, and analytical properties of the path integral~\cite{Weingarten:1983uj,Vafa:1983tf} provide stringent constraints on the IR structure.
Further insight on the dynamics has been obtained from non-perturbative approaches such as Monte Carlo lattice simulations~\cite{Wilson:1974sk,Kogut:1982ds}
and functional methods such as the functional renormalisation group (fRG) \cite{Wetterich:1992yh,Morris:1993qb} and Schwinger-Dyson equations (SDE) \cite{Dyson:1949ha,Schwinger:1951ex}.

However, for purely chiral theories, where left- and right-handed fermions transform under distinct representations of the gauge group, most tools developed for vector-like systems do not directly apply. In particular, lattice simulations cannot currently realise genuinely chiral gauge theories without introducing doublers~\cite{Nielsen:1980rz,Nielsen:1981hk,Nielsen:1981xu}. Moreover, anomaly matching does not sufficiently constrain the structure of the confining IR phase. As a result, the IR dynamics remain largely unknown and a variety of scenarios have been proposed, ranging from symmetry-preserving phases to sequential symmetry breaking patterns. 

Chiral gauge theories have also been widely discussed in the context of physics beyond the Standard Model, including grand unified theories and preon models of quark and lepton substructure~\cite{Pati:1975md,Weinberg:1975gm,Terazawa:1976xx,Susskind:1978ms,Eichten:1979ah,Raby:1979my,Harari:1979gi,Kaplan:1983fs,Kaplan:1983sm,Cacciapaglia:2019vce,Bars:1981se}. Understanding their non-perturbative dynamics is therefore not only of theoretical interest but also of phenomenological relevance.

In the present work we tackle the IR limit of chiral gauge theories using the functional renormalisation group (fRG), which allows for a direct treatment of chiral fermions. We employ the minimal truncation of the effective action required to diagnose the emergence of colour confinement and dynamical symmetry breaking ($\dSSB$) in the fermion sector. This framework, originally developed in the context of QCD and related theories, has shown qualitative and semi-quantitative reliability in QCD-like systems with few flavours~\cite{Goertz:2024dnz}, where extensive results from functional and Monte Carlo approaches exist.

Building on recent developments in Georgi--Glashow theories~\cite{Li:2025tvu}, we extend the analysis to the Bars--Yankielowicz (BY)~\cite{Bars:1981se} class by including the effects of colour confinement. This class consists of $SU(N_c)$ gauge theories with two species of left-handed Weyl fermions: one $\chi$ in the two-index symmetric representation of the gauge group and $(N_c+4)$  $\psi$ in the anti-fundamental representation. Gauge-anomaly cancellation fixes the fermion multiplicities and implies a flavour symmetry $SU(N_c+4)\times U(1)$, where the $U(1)$ corresponds to the anomaly-free combination of the two independent fermion-number of $\chi$ and $\psi$. 

These theories are asymptotically free for any $N_c$ and are expected to either confine or possibly flow to a conformal phase. Assuming confinement, it was noted in~\cite{Bars:1981se} that baryonic bound states of the form $\langle \chi \psi \psi\rangle$ can fully saturate the ultraviolet (UV) flavour anomalies. As a consequence, and in sharp contrast to QCD, the IR spectrum could consist solely of massless baryons without accompanying Goldstone bosons. Such conjectured scenario has been explored in preon models~\cite{Cacciapaglia:2019vce,Bars:1981se} and recently appeared as a central ingredient in scenarios of symmetric mass generation~\cite{Tong:2021phe,Razamat:2020kyf,Wang:2022ucy,You:2017ltx}. In addition, these theories have been recently investigated using higher-form symmetries~\cite{Bolognesi:2022beq,Bolognesi:2021jzs,Bolognesi:2021hmg,Bolognesi:2021yni,Bolognesi:2020mpe,Bolognesi:2023xxv,Bolognesi:2024bnm,Bolognesi:2025vkb} and anomaly-mediated supersymmetry breaking~\cite{Csaki:2021aqv}. Nevertheless, no conclusive picture of their IR dynamics is present.

The functional approach provides well-defined criteria for colour confinement and $\dSSB$, allowing us to investigate, for the first time, the IR limit while accounting for the interplay of these mechanisms. This work therefore provides a starting point for systematic studies of general chiral gauge theories and their novel dynamical regimes from first principles.

\section*{Functional approach to gauge-fermion theories}\label{sec:fRGapproach}
%
%\textbf{\textit{Functional renormalisation group approach to gauge-fermion theories.}}
The effective action formalism provides a general framework for deriving dressed correlation functions which encode the information on non-perturbative phenomena such as confinement and $\dSSB$. Within the fRG, quantum corrections are included through an IR regulator function ($R_k$) that suppresses momentum modes in the path integral measure below a cutoff scale $k$. This insertion allows introducing a scale-dependent effective action $\Gamma_k[\Phi]$ which interpolates between the microscopic classical action, and the full quantum effective action at $k=0$, where the regulator naturally vanishes. Its evolution over $k$ is governed by the flow equation~\cite{Wetterich:1992yh},
\begin{align}
k\partial_k\, \Gamma_k[\Phi] =\frac{1}{2} \mathrm{STr} \left[ \left(\Gamma_k^{(2)}[\Phi] + R_k\right)^{-1} k \partial_k R_k\right]\,,
\label{eq:wetterich}
\end{align}
where $\Gamma_k^{(n)}$ denotes the $n$th functional derivative of the effective action with respect to the fields $\Phi$ and $ \mathrm{STr}$ stands for the supertrace.

Over the past three decades, the fRG has been applied to a broad range of strongly coupled systems and non-perturbative dynamics from condensed matter, cold atoms, QCD to quantum gravity, see \cite{Dupuis:2020fhh} for a review. 
In gauge-fermion theories such as QCD, this program has proven highly successful, allowing to address the dynamical emergence of colour confinement and chiral symmetry breaking at a quantitative level even at finite temperature and chemical potential~\cite{Fu:2019hdw,Ihssen:2024miv,Mitter:2014wpa,Cyrol:2016tym,Cyrol:2017ewj,Cyrol:2017qkl,Gao:2020fbl,Gao:2020qsj,Fu:2022gou}. In the present work, we build on the existing technology and apply it to the novel context of chiral gauge theories.

Despite its compact and exact form, the flow equation can only be solved exactly for specific systems, so truncations or approximate solutions are generally necessary. Controlled expansion schemes exist which allow one to systematically enlarge the operator basis and monitor regulator dependences to assess reliability. In the fRG approach to gauge–fermion theories, a vertex expansion in operators is commonly employed, where the full effective action is reconstructed by solving a coupled set of flow equations for a truncated tower of operators. In this work, we include all classical tensor structures along with a Fierz-complete basis of four-fermion interactions.
This constitutes the minimal truncation so far known necessary to diagnose confinement and $\dSSB$. The effective action for the theories studied here is a functional of the classical fields which can be separated, without loss of generality, into three sectors,
\begin{align}\label{eq:fulleffectiveaction}
    &\Gamma_{k}[A_\mu,\bar c, c, \chi^\dagger, \chi, \psi^\dagger, \psi]= \Gamma_{{\rm gauge},\,k}[A_\mu,\bar c, c] \,+\,\\[.75ex]
    &\Gamma_{{{\rm gauge}{-}{\rm fermion},k}}[A_\mu, \chi^\dagger, \chi, \psi^\dagger, \psi] + \Gamma_{{\rm fermion},k} [\chi^\dagger, \chi, \psi^\dagger, \psi]\,.\notag
\end{align}
The first term includes all pure gauge interactions as well as the gauge-fixing and ghost terms necessary in gauge-fixed approaches. The last two terms account for the gauge-fermion and purely fermionic interactions.\\

\textbf{\textit{Colour confinement.}}
Hallmarks of colour confinement include the absence of colour-charged asymptotic states, the Wilson’s area-law scaling, and a vanishing Polyakov loop. These go hand in hand with the dynamical generation of a mass gap. Although the precise microscopic mechanism remains unknown, it is encoded in the pure gauge sector of the effective action and can be diagnosed at the level of correlation functions.

In covariant gauges, confinement can be characterised by the Kugo--Ojima criterion~\cite{Kugo:1979gm,Nakanishi:1990qm,Kugo:1995km} which establishes sufficient conditions for relating the existence of global BRST symmetry to colour-singlet states in the physical Hilbert space of a Yang-Mills theory.
These conditions impose a particular IR momentum scaling to the gluon and ghost propagators. In the Landau gauge, $Z_A \propto (p^2)^{-2\kappa}$ and $Z_c \propto (p^2)^{\kappa}$, implying an IR-suppressed gluon and an enhanced ghost, and thereby providing a unique IR closure for confining correlation functions~\cite{Kugo:1995km,vonSmekal:1997ohs,Alkofer:2000wg,Cyrol:2016tym,Zwanziger:2001kw,Lerche:2002ep}. In non-perturbative functional approaches, this allows one to bootstrap the confining solution without explicitly implementing a particular dynamical mechanism such as the Schwinger~\cite{Aguilar:2011xe,Aguilar:2021uwa,Aguilar:2022thg,Ferreira:2025anh} or quartet mechanisms~\cite{Alkofer:2011pe,Alkofer:2000wg}.

Within the fRG, confining correlators are obtained from the flow of the scale-dependent effective action, commonly in the Landau gauge. This includes the operators associated to $Z_A$ and $Z_c$, as well as to pure gauge, gauge--ghost, and gauge--fermion interactions. The relevance of the different $n$-point functions in the off-shell dynamics can be parametrised by exchange couplings
\begin{align}\label{eq:gaugeavatars}
&\alpha_{A^3} = {\left[\tilde\Gamma^{(A^3)}_k\right]^2}/{(4\pi\,Z_A^{3})} 
\,\,=\,\,\,\,\raisebox{-.4\height}{\includegraphics[width=2.cm]{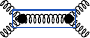}}\,
\,,\\[0.75ex]
&\alpha_{A^4} = {\tilde\Gamma^{(A^4)}_k}/{(4\pi\,Z_A^2)}\,\,\, {\rm and } \,\,\,  \alpha_{A\bar c c }  ={\left[\tilde\Gamma^{(A\bar c c )}_k\right]^2}/{(4\pi\,Z_c^2 Z_A)}\,,&\notag
\end{align}
where $\tilde\Gamma^{(n)}_k$ is the dressing of the respective vertices.
Equivalently, the strength of the gauge--fermion interactions can be parametrised by
\begin{align}\label{eq:gauge-fermionavatars}
&\alpha_{A\chi^\dagger\chi} = {\left[\tilde\Gamma^{(A\chi^\dagger\chi)}_k\right]^2}/{(4\pi\,Z_\chi^2 Z_A)}\,, 
&&\qquad{\rm and}\notag\\[1ex]
&\alpha_{A\psi^\dagger\psi} = {\left[\tilde\Gamma^{(A\psi^\dagger\psi)}_k\right]^2}/{(4\pi\,Z_\psi^2 Z_A)}\,.
\end{align}

Traditionally, resolving momentum dependencies of correlation functions requires substantial numerical effort, as the flow in $k$ must be computed for different external momenta, see e.g.~\cite{Cyrol:2016tym,Fu:2025hcm}. Here, we employ the method introduced in~\cite{Goertz:2024dnz}, which grants access to confining correlation functions directly at the level of the IR cutoff scale $k$. This reduces the computation to the vanishing external momentum limit, where the flow equations become significantly simpler and with certain regulator choices, analytical. The key ingredient is a convenient parametrisation of the gauge wave-function that incorporates the dynamically generated mass gap. See \cite{Goertz:2024dnz} and the supplemental material for  details. 

The introduction in the fRG of an IR regulator modifies the Slavnov--Taylor identities (STIs) for gauge invariance, leading to modified STIs (mSTIs) that reduce to the original as $k \to 0$. 
The mSTIs allow for a generation of a gauge mass gap, $\bar m_{\rm gap}^2$, along the RG flow which is both, regulator-induced and carries the underlying mechanism to confinement. While the former contribution drops in the $k\to0$ limit, correlation functions at the cutoff level do not necessarily reproduce the dependence on the external physical momentum. Therefore, within the current implementation, deriving confining correlators at the level of the cutoff requires minimising deviations from the STIs at $k>0$. This supposes the main source of uncertainty here and is controlled through the onset of power-law corrections to the IR scaling of the gauge mass gap. This is studied in detail in the supplemental material and we note that this discussion does not qualitatively impact the results here presented. Furthermore, this is  straightforwardly improved through the computation of momentum dependencies which is within the scope of a sequel to this work.

Furthermore, $\bar m_{\rm gap}^2$ is uniquely set by the Kugo--Ojima IR scaling and does not introduce any further freedom. This stands at the core of the bootstrap~\cite{Fischer:2008uz,Alkofer:2000wg,Cyrol:2016tym,Pawlowski:2003hq,Ferreira:2025tzo} approach to confinement and its further discussed in the supplemental material.\\

\begin{figure*}[th!]
	\centering
    \includegraphics[width=.6\columnwidth]{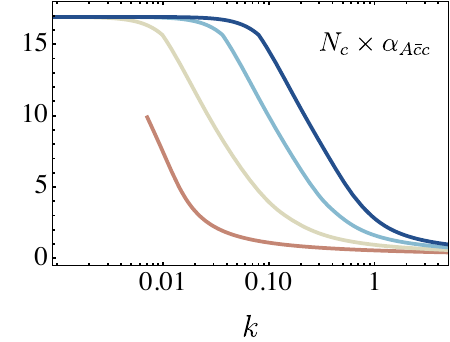}\hspace{.75cm}
    \includegraphics[width=.6\columnwidth]{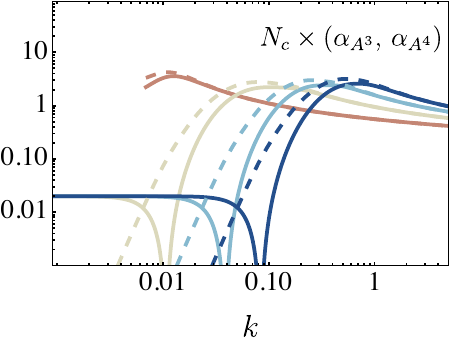}\hspace{.75cm}
    \includegraphics[width=.6\columnwidth]{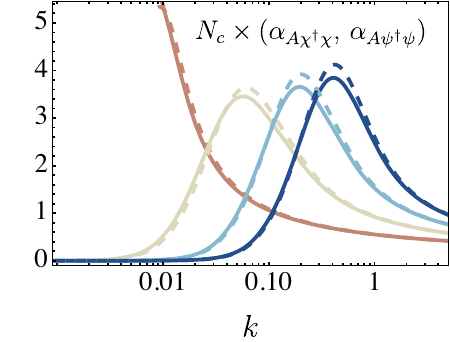}
	\caption{
    Ghost–gauge (left panel), three-gauge (centre panel, solid line),  four-gauge (centre panel, dashed line) gauge-$\chi$ (right panel, plain line) and gauge-$\psi$ (right panel, dashed line) exchange couplings for a BY theory with $N_c=3$ (red line), 4 (yellow), 5 (light blue) and 6 (dark blue).
    While for $N_c \geq 4$, the scaling confining solutions are shown, for $N_c = 3$, $\dSSB$ occurs, signalled by a singularity in the four-fermion couplings that propagates to the rest of the couplings. In this case, a nearby trajectory to the confining is displayed for illustration purposes.
    }
	\label{fig:alpha}
\end{figure*}
\textbf{\textit{Dynamical symmetry breaking.}}
The emergence of fermion bound states and condensates has been intensively studied with the effective action formalism, in particular in condensed matter and nuclear physics~\cite{Nambu:1961fr}. This dynamics is encoded in fermionic $n$-point functions and, via a Hubbard--Stratonovich transformation~\cite{HubbardPhysRevLett.3.77,Stratonovich} to bosonic formulation, it can be shown that a singularity in the dressings signals the transition from a symmetric to a broken phase~\cite{Miransky1989b,Miransky:1994vk,Gies:2005as,Braun:2011pp}.
Therefore, the appearance of such divergences provides a direct and unambiguous signal of $\dSSB$ as well as information about the characteristic scale of the phase transition and the condensate quantum numbers.

Fermion self-interactions are UV-irrelevant operators generated and partially driven by the gauge-fermion couplings. As these strengthen, the respective fermionic dressings grow and potentially diverge, turning relevant in the IR dynamics. To account for the complete dynamics in all symmetry channels and the formation of any possible bilinear condensate, it is important to consider a Fierz-complete four-fermion basis. For the kind of chiral gauge theories studied here this was derived in \cite{Li:2025tvu} and leads to the part of the effective action 
\begin{align}\label{eq:four-fermioneff}
    \Gamma_{{\rm fermion},k} [\chi^\dagger, \chi, \psi^\dagger, \psi] =&\\[1ex]
    &\hspace{-3.45cm} -\int_x  Z_{\psi}^2 \,  \sum_{i=1}^2  \lambda_{i} {\cal O}^\psi_i +Z_{\chi}^2 \sum_{i=4}^5  \lambda_i {\cal O}^\chi_i+Z_{\psi} Z_{\chi}\sum_{i=6}^7  \lambda_{i} {\cal O}^{\chi \psi}_i,\notag
\end{align}
with $\bar \lambda_i =k^2 \lambda_i$ being the dimensionless and scale-dependent dressings and $Z_{\psi,\,\chi}$ the respective fermion wave functions. The explicit operators are provided in the supplemental material.

The evolution of the four-fermion operators, and thus the IR fate of the symmetries, is governed by its RG flows. These have the form
\begin{align}\label{eq:4fflow}
    k\partial_k \bar \lambda_{i} =
   2\,\bar \lambda_{i}
   + \boldsymbol{c}^{\rm A}_{i}\, \alpha_g^2
   + \boldsymbol{c}^{\rm B}_{ij}\, \alpha_g\,\bar\lambda_{j}
   + \boldsymbol{c}^{\rm C }_{ijl}\, \bar\lambda_{j} \bar\lambda_{l}
   +\ldots\,\,\,,
\end{align}
where $\alpha_g$ denotes the appropriate gauge--fermion avatars in \eqref{eq:gauge-fermionavatars} and the coefficients $\boldsymbol{c}$ encode symmetry and combinatorial factors arising from colour, spin, and flavour traces of the respective loop diagrams. 
The first term in \eqref{eq:4fflow} accounts for the canonical scaling of the operator and the second term originates from gauge-mediated box diagrams which seed the flow near the asymptotically free UV fixed point. The remaining contributions describe mixed gauge--four-fermion diagrams and pure four-fermion interactions, which modify the operator scaling and induce mixing between channels. Altogether, the quadratic structure of the flow encodes the feedback of quantum fluctuations onto the couplings themselves, allowing for resonant growth and divergences. 

In the perturbative regime, where $\alpha_g\ll 1$, \cref{eq:4fflow} admits two fixed-point solutions: a near-Gaussian IR-attractive and IR-repulsive one. 
As the theory flows away from the free UV, the growth of the four-fermion couplings is initially bounded by the former fixed point. The subsequent evolution depends on the coefficients $\boldsymbol{c}$ which carry all the information about the symmetries of the theory. Depending on their sign and magnitude, both fixed points may approach and merge, unbounding the growth and allowing for a divergence inherent to $\dSSB$. 
We therefore define the critical coupling $\acrit$ necessary for $\dSSB$ as the minimum gauge coupling strength to induce such fixed-point merger.

The onset of $\dSSB$ in a purely fermionic formulation appears as a singularity and prevents the continuation of the flow into the deep IR. This can be conveniently addressed by an exact, scale-dependent field transformation into the bosonic formulation.
The procedure is implemented through the generalised flow equation for field redefinitions~\cite{Gies:2002hq,Pawlowski:2005xe}, which allows for the automatic inclusion of higher-dimensional operators. This improvement enables a smooth continuation of the RG flow into the deep IR as well as it improves the truncation of the effective action by effectively incorporating higher-dimensional fermionic operators.
For further details and discussions on $\dSSB$ and its implementations, see e.g.~\cite{Fu:2019hdw,Braun:2014ata,Goertz:2024dnz,Li:2025tvu,Braun:2011pp}.

\section*{Phase~structure}\label{sec:BYphasestructure}
To study the dynamics and access the IR of chiral gauge theories, we solve the coupled system of flow equations for $\{\alpha_i,\, \bar m_{\rm gap}^2,\, \bar \lambda_i,\,Z_i\}$ while imposing the Kugo--Ojima scaling solution in the IR. Within the present truncation, where dynamical bosonisation is not implemented, the flow can be continued to the deep IR only if $\dSSB$ does not occur. This requires that the leading gauge dynamics do not reach the critical strength $\acrit$ necessary to trigger a divergence in the four-fermion couplings. 

In the BY class, the ${\cal O}^\chi_4$ and ${\cal O}^\chi_5$ channels dominantly lead the fermionic dynamics and thus determine the onset of a potential singularity and the magnitude of $\acrit$.
Both operators are purely composed of $\chi$ fields and mix substantially in the sense that the divergence of the flow depends on the interplay of the two operators.
However, a rotation that diagonalizes the coefficient ${\boldsymbol{c}}^{\rm B}_{ij}$ in their respective flows provides a better basis in which a single operator, dominates and provides a better grip on the quantum numbers of the formed $\langle\chi\chi\rangle$ condensate.
Furthermore, the structure its flow equation allows for the fixed point merger  with the increasing gauge dynamics if ${\boldsymbol{c}^{\rm A}_i}/{\boldsymbol{c}^{\rm C}_{iii}}>0$. This is fulfilled only for $N_c\lesssim 5.30$ as the magnitude of $\boldsymbol{c}^{\rm A}_i$ vanishes causing an increment in $\acrit$ for larger $N_c$.
For further details on the derivation and analysis, see the supplemental material.

On the other hand, the confining gauge dynamics dictate the growth of the gauge-fermion couplings which consequently control the onset of $\dSSB$. In \Cref{fig:alpha}, we show the various gauge couplings defined in \eqref{eq:gaugeavatars} and \eqref{eq:gauge-fermionavatars} for different $N_c$.  These coincide in the perturbative UV, as imposed by the STI, and grow large  towards the IR. At around the confinement scale, the gauge propagator peaks, signalling the emergence of a mass gap for the gauge modes, namely confinement. Such an effect equivalently appears in the exchange gauge couplings, except for $\alpha_{A\bar c c }$. This coupling grows as the ghost contributions remain ungapped and freezes into constant value. Together with the zero-crossing and freezing of $\alpha_{A^3}$ these constitute hallmarks of confinement in the Landau gauge correlations in agreement with quantitative results from fRG \cite{Cyrol:2016tym,Fu:2025hcm,Ihssen:2024miv}, SDE \cite{Huber:2020keu,Huber:2018ned,Fischer:2002hna,Eichmann:2014xya,Papavassiliou:2022umz} and lattice simulations \cite{Athenodorou:2016oyh,Pinto-Gomez:2024mrk,Brito:2024aod,Ilgenfritz:2006he}. 

\begin{figure}
    \centering
    \includegraphics[width=0.475\columnwidth]{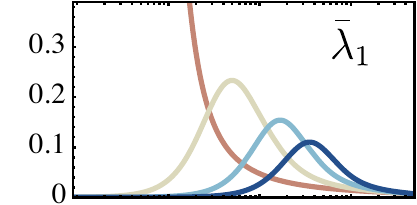}
    \includegraphics[width=0.475\columnwidth]{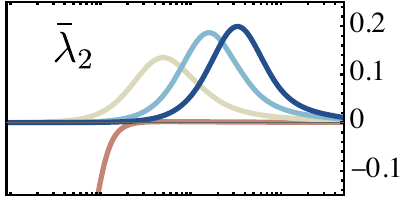}
    \includegraphics[width=0.475\columnwidth]{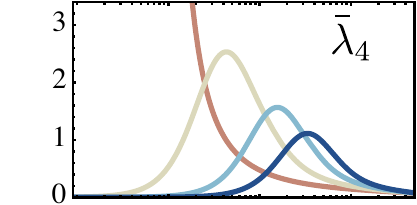}
    \includegraphics[width=0.475\columnwidth]{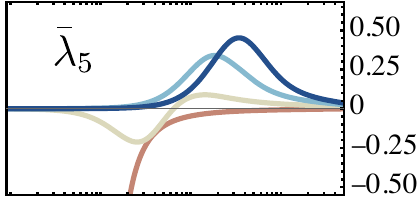}
    \includegraphics[width=0.475\columnwidth]{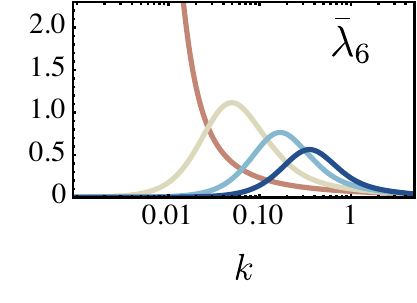}
    \includegraphics[width=0.475\columnwidth]{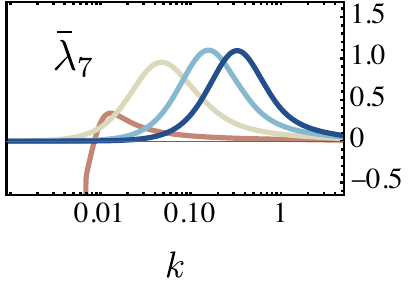}
    \caption{Four-fermion couplings, as defined in \eqref{eq:four-fermioneff}, for $N_c=3$ (red line), 4 (yellow), 5 (light blue) and 6 (dark blue) BY theories. }
    \label{fig:4fermionflows}
\end{figure}

In particular, $\alpha_{A\chi^\dagger\chi}$ controls the evolution of the dominant fermionic channel and is therefore the most relevant coupling for $\dSSB$. The integrated four-fermion couplings are shown in \Cref{fig:4fermionflows} for different values of $N_c$. Starting from the free UV, they grow as the gauge dynamics strengthen. For $N_c=3$, a divergence appears once the critical strength is exceeded, signalling $\dSSB$. In contrast, for $N_c=4,\,5,\,6$, the four-fermion dressings decrease below the confinement scale. In this case the gauge dynamics are not strong enough to trigger the fixed-point merger, and the flows are instead gapped by the emergence of the confining mass gap.

\begin{figure}[t!]
	\centering
	\includegraphics[width=.9\columnwidth]{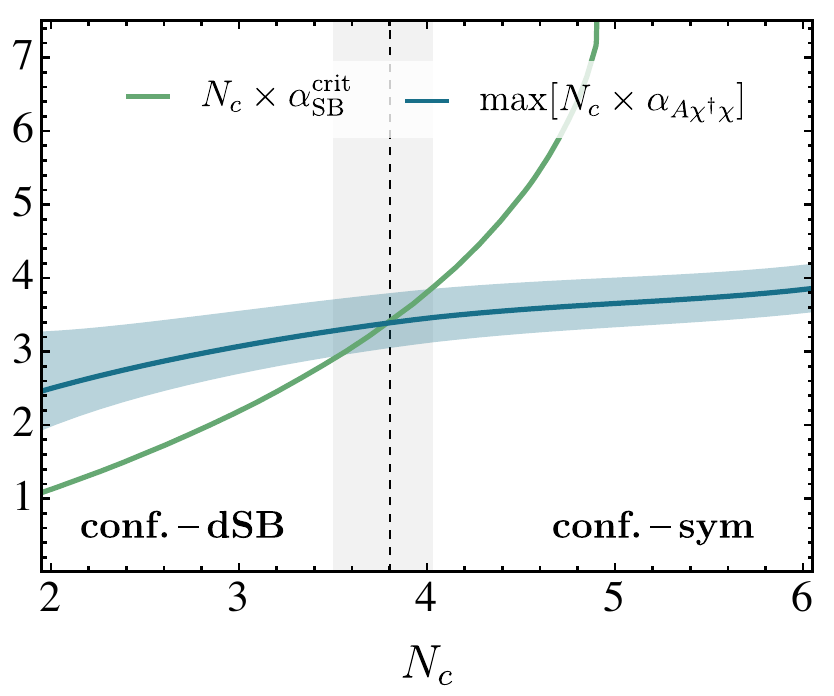}
	\caption{ $\acrit$ (green line) and the maximum value of the gauge-$\chi$ exchange coupling (blue line) as a function of $N_c$. As shaded blue region the estimated uncertainty in the minimisation of the STI. The vertical dashed line marks $N_c^{\rm crit}$ in \labelcref{eq:Nccrit}.}
	\label{fig:alphacrit}
\end{figure}

Consequently, two distinct phases arise, as shown in \Cref{fig:alphacrit}, where $N_c\,\acrit$ and $N_c\,\mathrm{max}\!\left[\alpha_{A\chi^\dagger\chi}\right]$ are compared. This indicates a phase transition at
\begin{align}\label{eq:Nccrit}
N_c^{\rm crit}= 3.81^{+0.23}_{-0.31}\,.
\end{align}
Its origin lies in the fermionic symmetry structure. While the pure gauge dynamics scale with $N_c$ and thus largely fix $\mathrm{max}\!\left[\alpha_{A\chi^\dagger\chi}\right]$, the increase of $\acrit$ with $N_c$, driven by the symmetries of the four-fermion sector, leads to the loss of $\dSSB$.

The error estimate quoted in \labelcref{eq:Nccrit} and shown by the shaded band in \Cref{fig:alphacrit} reflects the uncertainty associated with the minimisation of deviations from the STIs, as discussed in the supplemental material. Moreover, in theories exhibiting $\dSSB$ the RG flow cannot be continued arbitrarily deep into the IR within the purely fermionic description. In these cases, the peak of the $\alpha_{A\chi^\dagger\chi}$ has been estimated both from neighbouring quasi-scaling trajectories and from flows integrated without the divergent couplings. All estimates lie within the shaded band. 

In the symmetry-breaking phase, we note that for $N_c=2$, the fermion representations render the theory effectively vector-like and make the observed condensate consistent with Weingarten inequalities~\cite{Weingarten:1983uj} and anomaly matching. 

In the symmetry-preserving phase, the existence of massless baryonic states is consistent with the anomaly-matching scenario~\cite{Bars:1981se,Dimopoulos:1980hn} and with expectations from large-$N_c$ analyses~\cite{Eichten:1985fs,Karasik:2022gve}. While our results provide the first explicit dynamical study supporting this scenario, it does not yet select a unique IR realisation, and alternative baryonic or topological realisations may also be possible. We will further explore this regime in future works.

\section*{Conclusions}\label{sec:conclusions}
We have studied the confining and symmetry breaking dynamics of Bars--Yankielowicz chiral gauge theories using the non-perturbative functional renormalisation group which provides well-defined criteria for each phenomena.

Resolving the RG flows, we show the emergence of two distinct IR phases: for few colours, the theories confine accompanied by $\dSSB$ with a diquark-like $\langle \chi\chi\rangle$ condensate. For $N_c \gtrsim N_c^{\rm crit}$, confinement persists while $\dSSB$ disappears. 
This unusual phase is particularly interesting as fermions remain ungapped at all scales and massless baryons are expected in the spectrum. Moreover, exotic dynamics such as symmetric mass generation~\cite{Tong:2021phe,You:2017ltx,Razamat:2020kyf,Wang:2022ucy} may arise in this phase. A detailed study of its properties is deferred to future work.

The employed functional approach allows for straightforward and systematic improvements. The implementation of dynamical bosonisation will allow flowing beyond the $\dSSB$ scale and the inclusion of higher-dimensional fermionic operators, which improve the truncation and incorporate emergent baryons. Furthermore, full momentum dependencies can be computed reducing the uncertainty in the derivation of confining correlation functions and providing more precise estimates. 

Chiral gauge theories have long posed a major challenge for quantum field theory approaches. This work clarified a path towards their study from first principles and opens a broad landscape of gauge--fermion theories, allowing for the exploration of novel and exotic dynamical phenomena.

\section*{Acknowledgements}
We thank Reinhard Alkofer, Joannis Papavassiliou and Jan M. Pawlowski for discussions and suggestions on the manuscript. 
APG is supported by the RIKEN Special Postdoctoral Researcher (SPDR) Program. 
SV is supported by the Fonds de la Recherche Scientifique de Belgique (FNRS) under the IISN convention 4.4517.08. 
H-.L.L. is supported by the National Science Foundation of China under Grants No.1250050417 and by the start-up funding of Sun Yat-Sen University under grant number 74130-12255013.

\bibliography{fRG_chiral.bib}

%merlin.mbs apsrev4-1.bst 2010-07-25 4.21a (PWD, AO, DPC) hacked
%Control: key (0)
%Control: author (0) dotless jnrlst
%Control: editor formatted (1) identically to author
%Control: production of article title (0) allowed
%Control: page (1) range
%Control: year (0) verbatim
%Control: production of eprint (0) enabled
\begin{thebibliography}{98}%
\makeatletter
\providecommand \@ifxundefined [1]{%
 \@ifx{#1\undefined}
}%
\providecommand \@ifnum [1]{%
 \ifnum #1\expandafter \@firstoftwo
 \else \expandafter \@secondoftwo
 \fi
}%
\providecommand \@ifx [1]{%
 \ifx #1\expandafter \@firstoftwo
 \else \expandafter \@secondoftwo
 \fi
}%
\providecommand \natexlab [1]{#1}%
\providecommand \enquote  [1]{``#1''}%
\providecommand \bibnamefont  [1]{#1}%
\providecommand \bibfnamefont [1]{#1}%
\providecommand \citenamefont [1]{#1}%
\providecommand \href@noop [0]{\@secondoftwo}%
\providecommand \href [0]{\begingroup \@sanitize@url \@href}%
\providecommand \@href[1]{\@@startlink{#1}\@@href}%
\providecommand \@@href[1]{\endgroup#1\@@endlink}%
\providecommand \@sanitize@url [0]{\catcode `\\12\catcode `\$12\catcode `\&12\catcode `\#12\catcode `\^12\catcode `\_12\catcode `\%12\relax}%
\providecommand \@@startlink[1]{}%
\providecommand \@@endlink[0]{}%
\providecommand \url  [0]{\begingroup\@sanitize@url \@url }%
\providecommand \@url [1]{\endgroup\@href {#1}{\urlprefix }}%
\providecommand \urlprefix  [0]{URL }%
\providecommand \Eprint [0]{\href }%
\providecommand \doibase [0]{http://dx.doi.org/}%
\providecommand \selectlanguage [0]{\@gobble}%
\providecommand \bibinfo  [0]{\@secondoftwo}%
\providecommand \bibfield  [0]{\@secondoftwo}%
\providecommand \translation [1]{[#1]}%
\providecommand \BibitemOpen [0]{}%
\providecommand \bibitemStop [0]{}%
\providecommand \bibitemNoStop [0]{.\EOS\space}%
\providecommand \EOS [0]{\spacefactor3000\relax}%
\providecommand \BibitemShut  [1]{\csname bibitem#1\endcsname}%
\let\auto@bib@innerbib\@empty
%</preamble>
\bibitem [{\citenamefont {'t~Hooft}(1980)}]{tHooft:1979rat}%
  \BibitemOpen
  \bibfield  {author} {\bibinfo {author} {\bibfnamefont {Gerard}\ \bibnamefont {'t~Hooft}},\ }\bibfield  {title} {\enquote {\bibinfo {title} {{Naturalness, chiral symmetry, and spontaneous chiral symmetry breaking}},}\ }\href {\doibase 10.1007/978-1-4684-7571-5\_9} {\bibfield  {journal} {\bibinfo  {journal} {NATO Sci. Ser. B}\ }\textbf {\bibinfo {volume} {59}},\ \bibinfo {pages} {135--157} (\bibinfo {year} {1980})}\BibitemShut {NoStop}%
\bibitem [{\citenamefont {Seiberg}\ and\ \citenamefont {Witten}(1994{\natexlab{a}})}]{Seiberg:1994rs}%
  \BibitemOpen
  \bibfield  {author} {\bibinfo {author} {\bibfnamefont {N.}~\bibnamefont {Seiberg}}\ and\ \bibinfo {author} {\bibfnamefont {Edward}\ \bibnamefont {Witten}},\ }\bibfield  {title} {\enquote {\bibinfo {title} {{Electric - magnetic duality, monopole condensation, and confinement in N=2 supersymmetric Yang-Mills theory}},}\ }\href {\doibase 10.1016/0550-3213(94)90124-4} {\bibfield  {journal} {\bibinfo  {journal} {Nucl. Phys. B}\ }\textbf {\bibinfo {volume} {426}},\ \bibinfo {pages} {19--52} (\bibinfo {year} {1994}{\natexlab{a}})},\ \bibinfo {note} {[Erratum: Nucl.Phys.B 430, 485--486 (1994)]},\ \Eprint {http://arxiv.org/abs/hep-th/9407087} {arXiv:hep-th/9407087} \BibitemShut {NoStop}%
\bibitem [{\citenamefont {Seiberg}\ and\ \citenamefont {Witten}(1994{\natexlab{b}})}]{Seiberg:1994aj}%
  \BibitemOpen
  \bibfield  {author} {\bibinfo {author} {\bibfnamefont {N.}~\bibnamefont {Seiberg}}\ and\ \bibinfo {author} {\bibfnamefont {Edward}\ \bibnamefont {Witten}},\ }\bibfield  {title} {\enquote {\bibinfo {title} {{Monopoles, duality and chiral symmetry breaking in N=2 supersymmetric QCD}},}\ }\href {\doibase 10.1016/0550-3213(94)90214-3} {\bibfield  {journal} {\bibinfo  {journal} {Nucl. Phys. B}\ }\textbf {\bibinfo {volume} {431}},\ \bibinfo {pages} {484--550} (\bibinfo {year} {1994}{\natexlab{b}})},\ \Eprint {http://arxiv.org/abs/hep-th/9408099} {arXiv:hep-th/9408099} \BibitemShut {NoStop}%
\bibitem [{\citenamefont {Seiberg}(1994)}]{Seiberg:1994bz}%
  \BibitemOpen
  \bibfield  {author} {\bibinfo {author} {\bibfnamefont {Nathan}\ \bibnamefont {Seiberg}},\ }\bibfield  {title} {\enquote {\bibinfo {title} {{Exact results on the space of vacua of four-dimensional SUSY gauge theories}},}\ }\href {\doibase 10.1103/PhysRevD.49.6857} {\bibfield  {journal} {\bibinfo  {journal} {Phys. Rev. D}\ }\textbf {\bibinfo {volume} {49}},\ \bibinfo {pages} {6857--6863} (\bibinfo {year} {1994})},\ \Eprint {http://arxiv.org/abs/hep-th/9402044} {arXiv:hep-th/9402044} \BibitemShut {NoStop}%
\bibitem [{\citenamefont {Seiberg}(1995)}]{Seiberg:1994pq}%
  \BibitemOpen
  \bibfield  {author} {\bibinfo {author} {\bibfnamefont {N.}~\bibnamefont {Seiberg}},\ }\bibfield  {title} {\enquote {\bibinfo {title} {{Electric - magnetic duality in supersymmetric nonAbelian gauge theories}},}\ }\href {\doibase 10.1016/0550-3213(94)00023-8} {\bibfield  {journal} {\bibinfo  {journal} {Nucl. Phys. B}\ }\textbf {\bibinfo {volume} {435}},\ \bibinfo {pages} {129--146} (\bibinfo {year} {1995})},\ \Eprint {http://arxiv.org/abs/hep-th/9411149} {arXiv:hep-th/9411149} \BibitemShut {NoStop}%
\bibitem [{\citenamefont {Weingarten}(1983)}]{Weingarten:1983uj}%
  \BibitemOpen
  \bibfield  {author} {\bibinfo {author} {\bibfnamefont {Don}\ \bibnamefont {Weingarten}},\ }\bibfield  {title} {\enquote {\bibinfo {title} {{Mass Inequalities for QCD}},}\ }\href {\doibase 10.1103/PhysRevLett.51.1830} {\bibfield  {journal} {\bibinfo  {journal} {Phys. Rev. Lett.}\ }\textbf {\bibinfo {volume} {51}},\ \bibinfo {pages} {1830} (\bibinfo {year} {1983})}\BibitemShut {NoStop}%
\bibitem [{\citenamefont {Vafa}\ and\ \citenamefont {Witten}(1984)}]{Vafa:1983tf}%
  \BibitemOpen
  \bibfield  {author} {\bibinfo {author} {\bibfnamefont {C.}~\bibnamefont {Vafa}}\ and\ \bibinfo {author} {\bibfnamefont {Edward}\ \bibnamefont {Witten}},\ }\bibfield  {title} {\enquote {\bibinfo {title} {{Restrictions on Symmetry Breaking in Vector-Like Gauge Theories}},}\ }\href {\doibase 10.1016/0550-3213(84)90230-X} {\bibfield  {journal} {\bibinfo  {journal} {Nucl. Phys. B}\ }\textbf {\bibinfo {volume} {234}},\ \bibinfo {pages} {173--188} (\bibinfo {year} {1984})}\BibitemShut {NoStop}%
\bibitem [{\citenamefont {Wilson}(1974)}]{Wilson:1974sk}%
  \BibitemOpen
  \bibfield  {author} {\bibinfo {author} {\bibfnamefont {Kenneth~G.}\ \bibnamefont {Wilson}},\ }\bibfield  {title} {\enquote {\bibinfo {title} {{Confinement of Quarks}},}\ }\href {\doibase 10.1103/PhysRevD.10.2445} {\bibfield  {journal} {\bibinfo  {journal} {Phys. Rev. D}\ }\textbf {\bibinfo {volume} {10}},\ \bibinfo {pages} {2445--2459} (\bibinfo {year} {1974})}\BibitemShut {NoStop}%
\bibitem [{\citenamefont {Kogut}(1983)}]{Kogut:1982ds}%
  \BibitemOpen
  \bibfield  {author} {\bibinfo {author} {\bibfnamefont {John~B.}\ \bibnamefont {Kogut}},\ }\bibfield  {title} {\enquote {\bibinfo {title} {{A Review of the Lattice Gauge Theory Approach to Quantum Chromodynamics}},}\ }\href {\doibase 10.1103/RevModPhys.55.775} {\bibfield  {journal} {\bibinfo  {journal} {Rev. Mod. Phys.}\ }\textbf {\bibinfo {volume} {55}},\ \bibinfo {pages} {775} (\bibinfo {year} {1983})}\BibitemShut {NoStop}%
\bibitem [{\citenamefont {Wetterich}(1993)}]{Wetterich:1992yh}%
  \BibitemOpen
  \bibfield  {author} {\bibinfo {author} {\bibfnamefont {Christof}\ \bibnamefont {Wetterich}},\ }\bibfield  {title} {\enquote {\bibinfo {title} {{Exact evolution equation for the effective potential}},}\ }\href {\doibase 10.1016/0370-2693(93)90726-X} {\bibfield  {journal} {\bibinfo  {journal} {Phys. Lett.}\ }\textbf {\bibinfo {volume} {B301}},\ \bibinfo {pages} {90--94} (\bibinfo {year} {1993})},\ \Eprint {http://arxiv.org/abs/1710.05815} {arXiv:1710.05815 [hep-th]} \BibitemShut {NoStop}%
%\%CITATION = ARXIV:1710.05815;\%\%
\bibitem [{\citenamefont {Morris}(1994)}]{Morris:1993qb}%
  \BibitemOpen
  \bibfield  {author} {\bibinfo {author} {\bibfnamefont {Tim~R.}\ \bibnamefont {Morris}},\ }\bibfield  {title} {\enquote {\bibinfo {title} {{The Exact renormalization group and approximate solutions}},}\ }\href {\doibase 10.1142/S0217751X94000972} {\bibfield  {journal} {\bibinfo  {journal} {Int. J. Mod. Phys.}\ }\textbf {\bibinfo {volume} {A9}},\ \bibinfo {pages} {2411--2450} (\bibinfo {year} {1994})},\ \Eprint {http://arxiv.org/abs/hep-ph/9308265} {arXiv:hep-ph/9308265} \BibitemShut {NoStop}%
%\%CITATION = HEP-PH/9308265;\%\%
\bibitem [{\citenamefont {Dyson}(1949)}]{Dyson:1949ha}%
  \BibitemOpen
  \bibfield  {author} {\bibinfo {author} {\bibfnamefont {F.~J.}\ \bibnamefont {Dyson}},\ }\bibfield  {title} {\enquote {\bibinfo {title} {{The S matrix in quantum electrodynamics}},}\ }\href {\doibase 10.1103/PhysRev.75.1736} {\bibfield  {journal} {\bibinfo  {journal} {Phys. Rev.}\ }\textbf {\bibinfo {volume} {75}},\ \bibinfo {pages} {1736--1755} (\bibinfo {year} {1949})}\BibitemShut {NoStop}%
%\%CITATION = PHRVA,75,1736;\%\%
\bibitem [{\citenamefont {Schwinger}(1951)}]{Schwinger:1951ex}%
  \BibitemOpen
  \bibfield  {author} {\bibinfo {author} {\bibfnamefont {Julian~S.}\ \bibnamefont {Schwinger}},\ }\bibfield  {title} {\enquote {\bibinfo {title} {{On the Green's functions of quantized fields. 1.}}}\ }\href {\doibase 10.1073/pnas.37.7.452} {\bibfield  {journal} {\bibinfo  {journal} {Proc. Nat. Acad. Sci.}\ }\textbf {\bibinfo {volume} {37}},\ \bibinfo {pages} {452--455} (\bibinfo {year} {1951})}\BibitemShut {NoStop}%
%\%CITATION = PNASA,37,452;\%\%
\bibitem [{\citenamefont {Nielsen}\ and\ \citenamefont {Ninomiya}(1981{\natexlab{a}})}]{Nielsen:1980rz}%
  \BibitemOpen
  \bibfield  {author} {\bibinfo {author} {\bibfnamefont {Holger~Bech}\ \bibnamefont {Nielsen}}\ and\ \bibinfo {author} {\bibfnamefont {M.}~\bibnamefont {Ninomiya}},\ }\bibfield  {title} {\enquote {\bibinfo {title} {{Absence of Neutrinos on a Lattice. 1. Proof by Homotopy Theory}},}\ }\href {\doibase 10.1016/0550-3213(82)90011-6} {\bibfield  {journal} {\bibinfo  {journal} {Nucl. Phys. B}\ }\textbf {\bibinfo {volume} {185}},\ \bibinfo {pages} {20} (\bibinfo {year} {1981}{\natexlab{a}})},\ \bibinfo {note} {[Erratum: Nucl.Phys.B 195, 541 (1982)]}\BibitemShut {NoStop}%
\bibitem [{\citenamefont {Nielsen}\ and\ \citenamefont {Ninomiya}(1981{\natexlab{b}})}]{Nielsen:1981hk}%
  \BibitemOpen
  \bibfield  {author} {\bibinfo {author} {\bibfnamefont {Holger~Bech}\ \bibnamefont {Nielsen}}\ and\ \bibinfo {author} {\bibfnamefont {M.}~\bibnamefont {Ninomiya}},\ }\bibfield  {title} {\enquote {\bibinfo {title} {{No Go Theorem for Regularizing Chiral Fermions}},}\ }\href {\doibase 10.1016/0370-2693(81)91026-1} {\bibfield  {journal} {\bibinfo  {journal} {Phys. Lett. B}\ }\textbf {\bibinfo {volume} {105}},\ \bibinfo {pages} {219--223} (\bibinfo {year} {1981}{\natexlab{b}})}\BibitemShut {NoStop}%
\bibitem [{\citenamefont {Nielsen}\ and\ \citenamefont {Ninomiya}(1981{\natexlab{c}})}]{Nielsen:1981xu}%
  \BibitemOpen
  \bibfield  {author} {\bibinfo {author} {\bibfnamefont {Holger~Bech}\ \bibnamefont {Nielsen}}\ and\ \bibinfo {author} {\bibfnamefont {M.}~\bibnamefont {Ninomiya}},\ }\bibfield  {title} {\enquote {\bibinfo {title} {{Absence of Neutrinos on a Lattice. 2. Intuitive Topological Proof}},}\ }\href {\doibase 10.1016/0550-3213(81)90524-1} {\bibfield  {journal} {\bibinfo  {journal} {Nucl. Phys. B}\ }\textbf {\bibinfo {volume} {193}},\ \bibinfo {pages} {173--194} (\bibinfo {year} {1981}{\natexlab{c}})}\BibitemShut {NoStop}%
\bibitem [{\citenamefont {Pati}\ \emph {et~al.}(1975)\citenamefont {Pati}, \citenamefont {Salam},\ and\ \citenamefont {Strathdee}}]{Pati:1975md}%
  \BibitemOpen
  \bibfield  {author} {\bibinfo {author} {\bibfnamefont {Jogesh~C.}\ \bibnamefont {Pati}}, \bibinfo {author} {\bibfnamefont {Abdus}\ \bibnamefont {Salam}}, \ and\ \bibinfo {author} {\bibfnamefont {J.~A.}\ \bibnamefont {Strathdee}},\ }\bibfield  {title} {\enquote {\bibinfo {title} {{Are Quarks Composite?}}}\ }\href {\doibase 10.1016/0370-2693(75)90042-8} {\bibfield  {journal} {\bibinfo  {journal} {Phys. Lett. B}\ }\textbf {\bibinfo {volume} {59}},\ \bibinfo {pages} {265--268} (\bibinfo {year} {1975})}\BibitemShut {NoStop}%
\bibitem [{\citenamefont {Weinberg}(1976)}]{Weinberg:1975gm}%
  \BibitemOpen
  \bibfield  {author} {\bibinfo {author} {\bibfnamefont {Steven}\ \bibnamefont {Weinberg}},\ }\bibfield  {title} {\enquote {\bibinfo {title} {{Implications of Dynamical Symmetry Breaking}},}\ }\href {\doibase 10.1103/PhysRevD.19.1277} {\bibfield  {journal} {\bibinfo  {journal} {Phys. Rev. D}\ }\textbf {\bibinfo {volume} {13}},\ \bibinfo {pages} {974--996} (\bibinfo {year} {1976})},\ \bibinfo {note} {[Addendum: Phys.Rev.D 19, 1277--1280 (1979)]}\BibitemShut {NoStop}%
\bibitem [{\citenamefont {Terazawa}\ \emph {et~al.}(1977)\citenamefont {Terazawa}, \citenamefont {Akama},\ and\ \citenamefont {Chikashige}}]{Terazawa:1976xx}%
  \BibitemOpen
  \bibfield  {author} {\bibinfo {author} {\bibfnamefont {H.}~\bibnamefont {Terazawa}}, \bibinfo {author} {\bibfnamefont {K.}~\bibnamefont {Akama}}, \ and\ \bibinfo {author} {\bibfnamefont {Y.}~\bibnamefont {Chikashige}},\ }\bibfield  {title} {\enquote {\bibinfo {title} {{Unified Model of the Nambu-Jona-Lasinio Type for All Elementary Particle Forces}},}\ }\href {\doibase 10.1103/PhysRevD.15.480} {\bibfield  {journal} {\bibinfo  {journal} {Phys. Rev. D}\ }\textbf {\bibinfo {volume} {15}},\ \bibinfo {pages} {480} (\bibinfo {year} {1977})}\BibitemShut {NoStop}%
\bibitem [{\citenamefont {Susskind}(1979)}]{Susskind:1978ms}%
  \BibitemOpen
  \bibfield  {author} {\bibinfo {author} {\bibfnamefont {Leonard}\ \bibnamefont {Susskind}},\ }\bibfield  {title} {\enquote {\bibinfo {title} {Dynamics of spontaneous symmetry breaking in the weinberg-salam theory},}\ }\href {\doibase 10.1103/PhysRevD.20.2619} {\bibfield  {journal} {\bibinfo  {journal} {Phys. Rev. D}\ }\textbf {\bibinfo {volume} {20}},\ \bibinfo {pages} {2619--2625} (\bibinfo {year} {1979})}\BibitemShut {NoStop}%
\bibitem [{\citenamefont {Eichten}\ and\ \citenamefont {Lane}(1980)}]{Eichten:1979ah}%
  \BibitemOpen
  \bibfield  {author} {\bibinfo {author} {\bibfnamefont {Estia}\ \bibnamefont {Eichten}}\ and\ \bibinfo {author} {\bibfnamefont {Kenneth~D.}\ \bibnamefont {Lane}},\ }\bibfield  {title} {\enquote {\bibinfo {title} {{Dynamical Breaking of Weak Interaction Symmetries}},}\ }\href {\doibase 10.1016/0370-2693(80)90065-9} {\bibfield  {journal} {\bibinfo  {journal} {Phys. Lett. B}\ }\textbf {\bibinfo {volume} {90}},\ \bibinfo {pages} {125--130} (\bibinfo {year} {1980})}\BibitemShut {NoStop}%
\bibitem [{\citenamefont {Raby}\ \emph {et~al.}(1980)\citenamefont {Raby}, \citenamefont {Dimopoulos},\ and\ \citenamefont {Susskind}}]{Raby:1979my}%
  \BibitemOpen
  \bibfield  {author} {\bibinfo {author} {\bibfnamefont {Stuart}\ \bibnamefont {Raby}}, \bibinfo {author} {\bibfnamefont {Savas}\ \bibnamefont {Dimopoulos}}, \ and\ \bibinfo {author} {\bibfnamefont {Leonard}\ \bibnamefont {Susskind}},\ }\bibfield  {title} {\enquote {\bibinfo {title} {{Tumbling Gauge Theories}},}\ }\href {\doibase 10.1016/0550-3213(80)90093-0} {\bibfield  {journal} {\bibinfo  {journal} {Nucl. Phys. B}\ }\textbf {\bibinfo {volume} {169}},\ \bibinfo {pages} {373--383} (\bibinfo {year} {1980})}\BibitemShut {NoStop}%
\bibitem [{\citenamefont {Harari}(1979)}]{Harari:1979gi}%
  \BibitemOpen
  \bibfield  {author} {\bibinfo {author} {\bibfnamefont {Haim}\ \bibnamefont {Harari}},\ }\bibfield  {title} {\enquote {\bibinfo {title} {{A Schematic Model of Quarks and Leptons}},}\ }\href {\doibase 10.1016/0370-2693(79)90626-9} {\bibfield  {journal} {\bibinfo  {journal} {Phys. Lett. B}\ }\textbf {\bibinfo {volume} {86}},\ \bibinfo {pages} {83--86} (\bibinfo {year} {1979})}\BibitemShut {NoStop}%
\bibitem [{\citenamefont {Kaplan}\ and\ \citenamefont {Georgi}(1984)}]{Kaplan:1983fs}%
  \BibitemOpen
  \bibfield  {author} {\bibinfo {author} {\bibfnamefont {David~B.}\ \bibnamefont {Kaplan}}\ and\ \bibinfo {author} {\bibfnamefont {Howard}\ \bibnamefont {Georgi}},\ }\bibfield  {title} {\enquote {\bibinfo {title} {{SU(2) x U(1) Breaking by Vacuum Misalignment}},}\ }\href {\doibase 10.1016/0370-2693(84)91177-8} {\bibfield  {journal} {\bibinfo  {journal} {Phys. Lett. B}\ }\textbf {\bibinfo {volume} {136}},\ \bibinfo {pages} {183--186} (\bibinfo {year} {1984})}\BibitemShut {NoStop}%
\bibitem [{\citenamefont {Kaplan}\ \emph {et~al.}(1984)\citenamefont {Kaplan}, \citenamefont {Georgi},\ and\ \citenamefont {Dimopoulos}}]{Kaplan:1983sm}%
  \BibitemOpen
  \bibfield  {author} {\bibinfo {author} {\bibfnamefont {David~B}\ \bibnamefont {Kaplan}}, \bibinfo {author} {\bibfnamefont {Howard}\ \bibnamefont {Georgi}}, \ and\ \bibinfo {author} {\bibfnamefont {Savas}\ \bibnamefont {Dimopoulos}},\ }\bibfield  {title} {\enquote {\bibinfo {title} {Composite higgs scalars},}\ }\href {\doibase 10.1016/0370-2693(84)91178-X} {\bibfield  {journal} {\bibinfo  {journal} {Phys. Lett. B}\ }\textbf {\bibinfo {volume} {136B}},\ \bibinfo {pages} {187--190} (\bibinfo {year} {1984})}\BibitemShut {NoStop}%
\bibitem [{\citenamefont {Cacciapaglia}\ \emph {et~al.}(2019)\citenamefont {Cacciapaglia}, \citenamefont {Vatani},\ and\ \citenamefont {Wang}}]{Cacciapaglia:2019vce}%
  \BibitemOpen
  \bibfield  {author} {\bibinfo {author} {\bibfnamefont {Giacomo}\ \bibnamefont {Cacciapaglia}}, \bibinfo {author} {\bibfnamefont {Shahram}\ \bibnamefont {Vatani}}, \ and\ \bibinfo {author} {\bibfnamefont {Zhi-Wei}\ \bibnamefont {Wang}},\ }\bibfield  {title} {\enquote {\bibinfo {title} {{Tumbling to the Top}},}\ }\href@noop {} {\  (\bibinfo {year} {2019})},\ \Eprint {http://arxiv.org/abs/1909.08628} {arXiv:1909.08628 [hep-ph]} \BibitemShut {NoStop}%
\bibitem [{\citenamefont {Bars}\ and\ \citenamefont {Yankielowicz}(1981)}]{Bars:1981se}%
  \BibitemOpen
  \bibfield  {author} {\bibinfo {author} {\bibfnamefont {Itzhak}\ \bibnamefont {Bars}}\ and\ \bibinfo {author} {\bibfnamefont {Shimon}\ \bibnamefont {Yankielowicz}},\ }\bibfield  {title} {\enquote {\bibinfo {title} {{Composite Quarks and Leptons as Solutions of Anomaly Constraints}},}\ }\href {\doibase 10.1016/0370-2693(81)90664-X} {\bibfield  {journal} {\bibinfo  {journal} {Phys. Lett. B}\ }\textbf {\bibinfo {volume} {101}},\ \bibinfo {pages} {159--165} (\bibinfo {year} {1981})}\BibitemShut {NoStop}%
\bibitem [{\citenamefont {Goertz}\ \emph {et~al.}(2025)\citenamefont {Goertz}, \citenamefont {Pastor-Guti{\'e}rrez},\ and\ \citenamefont {Pawlowski}}]{Goertz:2024dnz}%
  \BibitemOpen
  \bibfield  {author} {\bibinfo {author} {\bibfnamefont {Florian}\ \bibnamefont {Goertz}}, \bibinfo {author} {\bibfnamefont {{\'A}lvaro}\ \bibnamefont {Pastor-Guti{\'e}rrez}}, \ and\ \bibinfo {author} {\bibfnamefont {Jan~M.}\ \bibnamefont {Pawlowski}},\ }\bibfield  {title} {\enquote {\bibinfo {title} {{Gauge-fermion cartography: From confinement and chiral symmetry breaking to conformality}},}\ }\href {\doibase 10.1103/7dzj-k6k8} {\bibfield  {journal} {\bibinfo  {journal} {Phys. Rev. D}\ }\textbf {\bibinfo {volume} {112}},\ \bibinfo {pages} {034029} (\bibinfo {year} {2025})},\ \Eprint {http://arxiv.org/abs/2412.12254} {arXiv:2412.12254 [hep-th]} \BibitemShut {NoStop}%
\bibitem [{\citenamefont {Li}\ \emph {et~al.}(2025)\citenamefont {Li}, \citenamefont {Pastor-Guti{\'e}rrez}, \citenamefont {Vatani},\ and\ \citenamefont {Xu}}]{Li:2025tvu}%
  \BibitemOpen
  \bibfield  {author} {\bibinfo {author} {\bibfnamefont {Hao-Lin}\ \bibnamefont {Li}}, \bibinfo {author} {\bibfnamefont {{\'A}lvaro}\ \bibnamefont {Pastor-Guti{\'e}rrez}}, \bibinfo {author} {\bibfnamefont {Shahram}\ \bibnamefont {Vatani}}, \ and\ \bibinfo {author} {\bibfnamefont {Ling-Xiao}\ \bibnamefont {Xu}},\ }\bibfield  {title} {\enquote {\bibinfo {title} {{Dynamical symmetry breaking in Georgi-Glashow chiral-gauge theories}},}\ }\href {\doibase 10.1007/JHEP12(2025)020} {\bibfield  {journal} {\bibinfo  {journal} {JHEP}\ }\textbf {\bibinfo {volume} {12}},\ \bibinfo {pages} {020} (\bibinfo {year} {2025})},\ \Eprint {http://arxiv.org/abs/2507.21208} {arXiv:2507.21208 [hep-th]} \BibitemShut {NoStop}%
\bibitem [{\citenamefont {Tong}(2022)}]{Tong:2021phe}%
  \BibitemOpen
  \bibfield  {author} {\bibinfo {author} {\bibfnamefont {David}\ \bibnamefont {Tong}},\ }\bibfield  {title} {\enquote {\bibinfo {title} {{Comments on symmetric mass generation in 2d and 4d}},}\ }\href {\doibase 10.1007/JHEP07(2022)001} {\bibfield  {journal} {\bibinfo  {journal} {JHEP}\ }\textbf {\bibinfo {volume} {07}},\ \bibinfo {pages} {001} (\bibinfo {year} {2022})},\ \Eprint {http://arxiv.org/abs/2104.03997} {arXiv:2104.03997 [hep-th]} \BibitemShut {NoStop}%
\bibitem [{\citenamefont {Razamat}\ and\ \citenamefont {Tong}(2021)}]{Razamat:2020kyf}%
  \BibitemOpen
  \bibfield  {author} {\bibinfo {author} {\bibfnamefont {Shlomo~S.}\ \bibnamefont {Razamat}}\ and\ \bibinfo {author} {\bibfnamefont {David}\ \bibnamefont {Tong}},\ }\bibfield  {title} {\enquote {\bibinfo {title} {{Gapped Chiral Fermions}},}\ }\href {\doibase 10.1103/PhysRevX.11.011063} {\bibfield  {journal} {\bibinfo  {journal} {Phys. Rev. X}\ }\textbf {\bibinfo {volume} {11}},\ \bibinfo {pages} {011063} (\bibinfo {year} {2021})},\ \Eprint {http://arxiv.org/abs/2009.05037} {arXiv:2009.05037 [hep-th]} \BibitemShut {NoStop}%
\bibitem [{\citenamefont {Wang}\ and\ \citenamefont {You}(2022)}]{Wang:2022ucy}%
  \BibitemOpen
  \bibfield  {author} {\bibinfo {author} {\bibfnamefont {Juven}\ \bibnamefont {Wang}}\ and\ \bibinfo {author} {\bibfnamefont {Yi-Zhuang}\ \bibnamefont {You}},\ }\bibfield  {title} {\enquote {\bibinfo {title} {{Symmetric Mass Generation}},}\ }\href {\doibase 10.3390/sym14071475} {\bibfield  {journal} {\bibinfo  {journal} {Symmetry}\ }\textbf {\bibinfo {volume} {14}},\ \bibinfo {pages} {1475} (\bibinfo {year} {2022})},\ \Eprint {http://arxiv.org/abs/2204.14271} {arXiv:2204.14271 [cond-mat.str-el]} \BibitemShut {NoStop}%
\bibitem [{\citenamefont {You}\ \emph {et~al.}(2018)\citenamefont {You}, \citenamefont {He}, \citenamefont {Xu},\ and\ \citenamefont {Vishwanath}}]{You:2017ltx}%
  \BibitemOpen
  \bibfield  {author} {\bibinfo {author} {\bibfnamefont {Yi-Zhuang}\ \bibnamefont {You}}, \bibinfo {author} {\bibfnamefont {Yin-Chen}\ \bibnamefont {He}}, \bibinfo {author} {\bibfnamefont {Cenke}\ \bibnamefont {Xu}}, \ and\ \bibinfo {author} {\bibfnamefont {Ashvin}\ \bibnamefont {Vishwanath}},\ }\bibfield  {title} {\enquote {\bibinfo {title} {{Symmetric Fermion Mass Generation as Deconfined Quantum Criticality}},}\ }\href {\doibase 10.1103/PhysRevX.8.011026} {\bibfield  {journal} {\bibinfo  {journal} {Phys. Rev. X}\ }\textbf {\bibinfo {volume} {8}},\ \bibinfo {pages} {011026} (\bibinfo {year} {2018})},\ \Eprint {http://arxiv.org/abs/1705.09313} {arXiv:1705.09313 [cond-mat.str-el]} \BibitemShut {NoStop}%
\bibitem [{\citenamefont {Bolognesi}\ \emph {et~al.}(2022{\natexlab{a}})\citenamefont {Bolognesi}, \citenamefont {Konishi},\ and\ \citenamefont {Luzio}}]{Bolognesi:2022beq}%
  \BibitemOpen
  \bibfield  {author} {\bibinfo {author} {\bibfnamefont {Stefano}\ \bibnamefont {Bolognesi}}, \bibinfo {author} {\bibfnamefont {Kenichi}\ \bibnamefont {Konishi}}, \ and\ \bibinfo {author} {\bibfnamefont {Andrea}\ \bibnamefont {Luzio}},\ }\bibfield  {title} {\enquote {\bibinfo {title} {{Dynamical Abelianization and anomalies in chiral gauge theories}},}\ }\href {\doibase 10.1007/JHEP12(2022)110} {\bibfield  {journal} {\bibinfo  {journal} {JHEP}\ }\textbf {\bibinfo {volume} {12}},\ \bibinfo {pages} {110} (\bibinfo {year} {2022}{\natexlab{a}})},\ \Eprint {http://arxiv.org/abs/2206.00538} {arXiv:2206.00538 [hep-th]} \BibitemShut {NoStop}%
\bibitem [{\citenamefont {Bolognesi}\ \emph {et~al.}(2022{\natexlab{b}})\citenamefont {Bolognesi}, \citenamefont {Konishi},\ and\ \citenamefont {Luzio}}]{Bolognesi:2021jzs}%
  \BibitemOpen
  \bibfield  {author} {\bibinfo {author} {\bibfnamefont {Stefano}\ \bibnamefont {Bolognesi}}, \bibinfo {author} {\bibfnamefont {Kenichi}\ \bibnamefont {Konishi}}, \ and\ \bibinfo {author} {\bibfnamefont {Andrea}\ \bibnamefont {Luzio}},\ }\bibfield  {title} {\enquote {\bibinfo {title} {{Anomalies and phases of strongly coupled chiral gauge theories: Recent developments}},}\ }\href {\doibase 10.1142/S0217751X22300149} {\bibfield  {journal} {\bibinfo  {journal} {Int. J. Mod. Phys. A}\ }\textbf {\bibinfo {volume} {37}},\ \bibinfo {pages} {2230014} (\bibinfo {year} {2022}{\natexlab{b}})},\ \Eprint {http://arxiv.org/abs/2110.02104} {arXiv:2110.02104 [hep-th]} \BibitemShut {NoStop}%
\bibitem [{\citenamefont {Bolognesi}\ \emph {et~al.}(2021{\natexlab{a}})\citenamefont {Bolognesi}, \citenamefont {Konishi},\ and\ \citenamefont {Luzio}}]{Bolognesi:2021hmg}%
  \BibitemOpen
  \bibfield  {author} {\bibinfo {author} {\bibfnamefont {Stefano}\ \bibnamefont {Bolognesi}}, \bibinfo {author} {\bibfnamefont {Kenichi}\ \bibnamefont {Konishi}}, \ and\ \bibinfo {author} {\bibfnamefont {Andrea}\ \bibnamefont {Luzio}},\ }\bibfield  {title} {\enquote {\bibinfo {title} {{Strong anomaly and phases of chiral gauge theories}},}\ }\href {\doibase 10.1007/JHEP08(2021)028} {\bibfield  {journal} {\bibinfo  {journal} {JHEP}\ }\textbf {\bibinfo {volume} {08}},\ \bibinfo {pages} {028} (\bibinfo {year} {2021}{\natexlab{a}})},\ \Eprint {http://arxiv.org/abs/2105.03921} {arXiv:2105.03921 [hep-th]} \BibitemShut {NoStop}%
\bibitem [{\citenamefont {Bolognesi}\ \emph {et~al.}(2021{\natexlab{b}})\citenamefont {Bolognesi}, \citenamefont {Konishi},\ and\ \citenamefont {Luzio}}]{Bolognesi:2021yni}%
  \BibitemOpen
  \bibfield  {author} {\bibinfo {author} {\bibfnamefont {Stefano}\ \bibnamefont {Bolognesi}}, \bibinfo {author} {\bibfnamefont {Kenichi}\ \bibnamefont {Konishi}}, \ and\ \bibinfo {author} {\bibfnamefont {Andrea}\ \bibnamefont {Luzio}},\ }\bibfield  {title} {\enquote {\bibinfo {title} {{Probing the dynamics of chiral $SU(N)$ gauge theories via generalized anomalies}},}\ }\href {\doibase 10.1103/PhysRevD.103.094016} {\bibfield  {journal} {\bibinfo  {journal} {Phys. Rev. D}\ }\textbf {\bibinfo {volume} {103}},\ \bibinfo {pages} {094016} (\bibinfo {year} {2021}{\natexlab{b}})},\ \Eprint {http://arxiv.org/abs/2101.02601} {arXiv:2101.02601 [hep-th]} \BibitemShut {NoStop}%
\bibitem [{\citenamefont {Bolognesi}\ \emph {et~al.}(2020)\citenamefont {Bolognesi}, \citenamefont {Konishi},\ and\ \citenamefont {Luzio}}]{Bolognesi:2020mpe}%
  \BibitemOpen
  \bibfield  {author} {\bibinfo {author} {\bibfnamefont {Stefano}\ \bibnamefont {Bolognesi}}, \bibinfo {author} {\bibfnamefont {Kenichi}\ \bibnamefont {Konishi}}, \ and\ \bibinfo {author} {\bibfnamefont {Andrea}\ \bibnamefont {Luzio}},\ }\bibfield  {title} {\enquote {\bibinfo {title} {{Dynamics from symmetries in chiral $SU(N)$ gauge theories}},}\ }\href {\doibase 10.1007/JHEP09(2020)001} {\bibfield  {journal} {\bibinfo  {journal} {JHEP}\ }\textbf {\bibinfo {volume} {09}},\ \bibinfo {pages} {001} (\bibinfo {year} {2020})},\ \Eprint {http://arxiv.org/abs/2004.06639} {arXiv:2004.06639 [hep-th]} \BibitemShut {NoStop}%
\bibitem [{\citenamefont {Bolognesi}\ \emph {et~al.}(2023)\citenamefont {Bolognesi}, \citenamefont {Konishi},\ and\ \citenamefont {Luzio}}]{Bolognesi:2023xxv}%
  \BibitemOpen
  \bibfield  {author} {\bibinfo {author} {\bibfnamefont {Stefano}\ \bibnamefont {Bolognesi}}, \bibinfo {author} {\bibfnamefont {Kenichi}\ \bibnamefont {Konishi}}, \ and\ \bibinfo {author} {\bibfnamefont {Andrea}\ \bibnamefont {Luzio}},\ }\bibfield  {title} {\enquote {\bibinfo {title} {{The \ensuremath{\mathbb{Z}}$_{2}$ anomaly in some chiral gauge theories}},}\ }\href {\doibase 10.1007/JHEP08(2023)125} {\bibfield  {journal} {\bibinfo  {journal} {JHEP}\ }\textbf {\bibinfo {volume} {08}},\ \bibinfo {pages} {125} (\bibinfo {year} {2023})},\ \Eprint {http://arxiv.org/abs/2307.03822} {arXiv:2307.03822 [hep-th]} \BibitemShut {NoStop}%
\bibitem [{\citenamefont {Bolognesi}\ \emph {et~al.}(2024)\citenamefont {Bolognesi}, \citenamefont {Konishi}, \citenamefont {Luzio},\ and\ \citenamefont {Orso}}]{Bolognesi:2024bnm}%
  \BibitemOpen
  \bibfield  {author} {\bibinfo {author} {\bibfnamefont {Stefano}\ \bibnamefont {Bolognesi}}, \bibinfo {author} {\bibfnamefont {Kenichi}\ \bibnamefont {Konishi}}, \bibinfo {author} {\bibfnamefont {Andrea}\ \bibnamefont {Luzio}}, \ and\ \bibinfo {author} {\bibfnamefont {Matteo}\ \bibnamefont {Orso}},\ }\bibfield  {title} {\enquote {\bibinfo {title} {{Natural Anomaly Matching}},}\ }\href@noop {} {\  (\bibinfo {year} {2024})},\ \Eprint {http://arxiv.org/abs/2410.01315} {arXiv:2410.01315 [hep-th]} \BibitemShut {NoStop}%
\bibitem [{\citenamefont {Bolognesi}\ \emph {et~al.}(2025)\citenamefont {Bolognesi}, \citenamefont {Luzio},\ and\ \citenamefont {Santoni}}]{Bolognesi:2025vkb}%
  \BibitemOpen
  \bibfield  {author} {\bibinfo {author} {\bibfnamefont {Stefano}\ \bibnamefont {Bolognesi}}, \bibinfo {author} {\bibfnamefont {Andrea}\ \bibnamefont {Luzio}}, \ and\ \bibinfo {author} {\bibfnamefont {Giacomo}\ \bibnamefont {Santoni}},\ }\bibfield  {title} {\enquote {\bibinfo {title} {{Baryons, Skyrmions and $θ$-periodicity anomaly in chiral and vector-like gauge theories}},}\ }\href@noop {} {\  (\bibinfo {year} {2025})},\ \Eprint {http://arxiv.org/abs/2510.09866} {arXiv:2510.09866 [hep-th]} \BibitemShut {NoStop}%
\bibitem [{\citenamefont {Cs\'aki}\ \emph {et~al.}(2022)\citenamefont {Cs\'aki}, \citenamefont {Murayama},\ and\ \citenamefont {Telem}}]{Csaki:2021aqv}%
  \BibitemOpen
  \bibfield  {author} {\bibinfo {author} {\bibfnamefont {Csaba}\ \bibnamefont {Cs\'aki}}, \bibinfo {author} {\bibfnamefont {Hitoshi}\ \bibnamefont {Murayama}}, \ and\ \bibinfo {author} {\bibfnamefont {Ofri}\ \bibnamefont {Telem}},\ }\bibfield  {title} {\enquote {\bibinfo {title} {{More exact results on chiral gauge theories: The case of the symmetric tensor}},}\ }\href {\doibase 10.1103/PhysRevD.105.045007} {\bibfield  {journal} {\bibinfo  {journal} {Phys. Rev. D}\ }\textbf {\bibinfo {volume} {105}},\ \bibinfo {pages} {045007} (\bibinfo {year} {2022})},\ \Eprint {http://arxiv.org/abs/2105.03444} {arXiv:2105.03444 [hep-th]} \BibitemShut {NoStop}%
\bibitem [{\citenamefont {Dupuis}\ \emph {et~al.}(2021)\citenamefont {Dupuis}, \citenamefont {Canet}, \citenamefont {Eichhorn}, \citenamefont {Metzner}, \citenamefont {Pawlowski}, \citenamefont {Tissier},\ and\ \citenamefont {Wschebor}}]{Dupuis:2020fhh}%
  \BibitemOpen
  \bibfield  {author} {\bibinfo {author} {\bibfnamefont {N.}~\bibnamefont {Dupuis}}, \bibinfo {author} {\bibfnamefont {L.}~\bibnamefont {Canet}}, \bibinfo {author} {\bibfnamefont {A.}~\bibnamefont {Eichhorn}}, \bibinfo {author} {\bibfnamefont {W.}~\bibnamefont {Metzner}}, \bibinfo {author} {\bibfnamefont {J.~M.}\ \bibnamefont {Pawlowski}}, \bibinfo {author} {\bibfnamefont {M.}~\bibnamefont {Tissier}}, \ and\ \bibinfo {author} {\bibfnamefont {N.}~\bibnamefont {Wschebor}},\ }\bibfield  {title} {\enquote {\bibinfo {title} {{The nonperturbative functional renormalization group and its applications}},}\ }\href {\doibase 10.1016/j.physrep.2021.01.001} {\bibfield  {journal} {\bibinfo  {journal} {Phys. Rept.}\ }\textbf {\bibinfo {volume} {910}},\ \bibinfo {pages} {1--114} (\bibinfo {year} {2021})},\ \Eprint {http://arxiv.org/abs/2006.04853} {arXiv:2006.04853 [cond-mat.stat-mech]} \BibitemShut {NoStop}%
\bibitem [{\citenamefont {Fu}\ \emph {et~al.}(2020)\citenamefont {Fu}, \citenamefont {Pawlowski},\ and\ \citenamefont {Rennecke}}]{Fu:2019hdw}%
  \BibitemOpen
  \bibfield  {author} {\bibinfo {author} {\bibfnamefont {Wei-jie}\ \bibnamefont {Fu}}, \bibinfo {author} {\bibfnamefont {Jan~M.}\ \bibnamefont {Pawlowski}}, \ and\ \bibinfo {author} {\bibfnamefont {Fabian}\ \bibnamefont {Rennecke}},\ }\bibfield  {title} {\enquote {\bibinfo {title} {Qcd phase structure at finite temperature and density},}\ }\href {\doibase 10.1103/PhysRevD.101.054032} {\bibfield  {journal} {\bibinfo  {journal} {Phys. Rev. D}\ }\textbf {\bibinfo {volume} {101}},\ \bibinfo {pages} {054032} (\bibinfo {year} {2020})},\ \Eprint {http://arxiv.org/abs/1909.02991} {arXiv:1909.02991 [hep-ph]} \BibitemShut {NoStop}%
\bibitem [{\citenamefont {Ihssen}\ \emph {et~al.}(2024)\citenamefont {Ihssen}, \citenamefont {Pawlowski}, \citenamefont {Sattler},\ and\ \citenamefont {Wink}}]{Ihssen:2024miv}%
  \BibitemOpen
  \bibfield  {author} {\bibinfo {author} {\bibfnamefont {Friederike}\ \bibnamefont {Ihssen}}, \bibinfo {author} {\bibfnamefont {Jan~M.}\ \bibnamefont {Pawlowski}}, \bibinfo {author} {\bibfnamefont {Franz~R.}\ \bibnamefont {Sattler}}, \ and\ \bibinfo {author} {\bibfnamefont {Nicolas}\ \bibnamefont {Wink}},\ }\bibfield  {title} {\enquote {\bibinfo {title} {Towards quantitative precision in functional qcd i},}\ }\href@noop {} {\  (\bibinfo {year} {2024})},\ \Eprint {http://arxiv.org/abs/2408.08413} {arXiv:2408.08413 [hep-ph]} \BibitemShut {NoStop}%
\bibitem [{\citenamefont {Mitter}\ \emph {et~al.}(2015)\citenamefont {Mitter}, \citenamefont {Pawlowski},\ and\ \citenamefont {Strodthoff}}]{Mitter:2014wpa}%
  \BibitemOpen
  \bibfield  {author} {\bibinfo {author} {\bibfnamefont {Mario}\ \bibnamefont {Mitter}}, \bibinfo {author} {\bibfnamefont {Jan~M.}\ \bibnamefont {Pawlowski}}, \ and\ \bibinfo {author} {\bibfnamefont {Nils}\ \bibnamefont {Strodthoff}},\ }\bibfield  {title} {\enquote {\bibinfo {title} {{Chiral symmetry breaking in continuum QCD}},}\ }\href {\doibase 10.1103/PhysRevD.91.054035} {\bibfield  {journal} {\bibinfo  {journal} {Phys. Rev.}\ }\textbf {\bibinfo {volume} {D91}},\ \bibinfo {pages} {054035} (\bibinfo {year} {2015})},\ \Eprint {http://arxiv.org/abs/1411.7978} {arXiv:1411.7978 [hep-ph]} \BibitemShut {NoStop}%
%\%CITATION = ARXIV:1411.7978;\%\%
\bibitem [{\citenamefont {Cyrol}\ \emph {et~al.}(2016)\citenamefont {Cyrol}, \citenamefont {Fister}, \citenamefont {Mitter}, \citenamefont {Pawlowski},\ and\ \citenamefont {Strodthoff}}]{Cyrol:2016tym}%
  \BibitemOpen
  \bibfield  {author} {\bibinfo {author} {\bibfnamefont {Anton~K.}\ \bibnamefont {Cyrol}}, \bibinfo {author} {\bibfnamefont {Leonard}\ \bibnamefont {Fister}}, \bibinfo {author} {\bibfnamefont {Mario}\ \bibnamefont {Mitter}}, \bibinfo {author} {\bibfnamefont {Jan~M.}\ \bibnamefont {Pawlowski}}, \ and\ \bibinfo {author} {\bibfnamefont {Nils}\ \bibnamefont {Strodthoff}},\ }\bibfield  {title} {\enquote {\bibinfo {title} {{Landau gauge Yang-Mills correlation functions}},}\ }\href {\doibase 10.1103/PhysRevD.94.054005} {\bibfield  {journal} {\bibinfo  {journal} {Phys. Rev.}\ }\textbf {\bibinfo {volume} {D94}},\ \bibinfo {pages} {054005} (\bibinfo {year} {2016})},\ \Eprint {http://arxiv.org/abs/1605.01856} {arXiv:1605.01856 [hep-ph]} \BibitemShut {NoStop}%
%\%CITATION = ARXIV:1605.01856;\%\%
\bibitem [{\citenamefont {Cyrol}\ \emph {et~al.}(2018{\natexlab{a}})\citenamefont {Cyrol}, \citenamefont {Mitter}, \citenamefont {Pawlowski},\ and\ \citenamefont {Strodthoff}}]{Cyrol:2017ewj}%
  \BibitemOpen
  \bibfield  {author} {\bibinfo {author} {\bibfnamefont {Anton~K.}\ \bibnamefont {Cyrol}}, \bibinfo {author} {\bibfnamefont {Mario}\ \bibnamefont {Mitter}}, \bibinfo {author} {\bibfnamefont {Jan~M.}\ \bibnamefont {Pawlowski}}, \ and\ \bibinfo {author} {\bibfnamefont {Nils}\ \bibnamefont {Strodthoff}},\ }\bibfield  {title} {\enquote {\bibinfo {title} {{Nonperturbative quark, gluon, and meson correlators of unquenched QCD}},}\ }\href {\doibase 10.1103/PhysRevD.97.054006} {\bibfield  {journal} {\bibinfo  {journal} {Phys. Rev.}\ }\textbf {\bibinfo {volume} {D97}},\ \bibinfo {pages} {054006} (\bibinfo {year} {2018}{\natexlab{a}})},\ \Eprint {http://arxiv.org/abs/1706.06326} {arXiv:1706.06326 [hep-ph]} \BibitemShut {NoStop}%
%\%CITATION = ARXIV:1706.06326;\%\%
\bibitem [{\citenamefont {Cyrol}\ \emph {et~al.}(2018{\natexlab{b}})\citenamefont {Cyrol}, \citenamefont {Mitter}, \citenamefont {Pawlowski},\ and\ \citenamefont {Strodthoff}}]{Cyrol:2017qkl}%
  \BibitemOpen
  \bibfield  {author} {\bibinfo {author} {\bibfnamefont {Anton~K.}\ \bibnamefont {Cyrol}}, \bibinfo {author} {\bibfnamefont {Mario}\ \bibnamefont {Mitter}}, \bibinfo {author} {\bibfnamefont {Jan~M.}\ \bibnamefont {Pawlowski}}, \ and\ \bibinfo {author} {\bibfnamefont {Nils}\ \bibnamefont {Strodthoff}},\ }\bibfield  {title} {\enquote {\bibinfo {title} {Nonperturbative finite-temperature yang-mills theory},}\ }\href {\doibase 10.1103/PhysRevD.97.054015} {\bibfield  {journal} {\bibinfo  {journal} {Phys. Rev. D}\ }\textbf {\bibinfo {volume} {97}},\ \bibinfo {pages} {054015} (\bibinfo {year} {2018}{\natexlab{b}})},\ \Eprint {http://arxiv.org/abs/1708.03482} {arXiv:1708.03482 [hep-ph]} \BibitemShut {NoStop}%
\bibitem [{\citenamefont {Gao}\ and\ \citenamefont {Pawlowski}(2021)}]{Gao:2020fbl}%
  \BibitemOpen
  \bibfield  {author} {\bibinfo {author} {\bibfnamefont {Fei}\ \bibnamefont {Gao}}\ and\ \bibinfo {author} {\bibfnamefont {Jan~M.}\ \bibnamefont {Pawlowski}},\ }\bibfield  {title} {\enquote {\bibinfo {title} {{Chiral phase structure and critical end point in QCD}},}\ }\href {\doibase 10.1016/j.physletb.2021.136584} {\bibfield  {journal} {\bibinfo  {journal} {Phys. Lett. B}\ }\textbf {\bibinfo {volume} {820}},\ \bibinfo {pages} {136584} (\bibinfo {year} {2021})},\ \Eprint {http://arxiv.org/abs/2010.13705} {arXiv:2010.13705 [hep-ph]} \BibitemShut {NoStop}%
\bibitem [{\citenamefont {Gao}\ and\ \citenamefont {Pawlowski}(2020)}]{Gao:2020qsj}%
  \BibitemOpen
  \bibfield  {author} {\bibinfo {author} {\bibfnamefont {Fei}\ \bibnamefont {Gao}}\ and\ \bibinfo {author} {\bibfnamefont {Jan~M.}\ \bibnamefont {Pawlowski}},\ }\bibfield  {title} {\enquote {\bibinfo {title} {{QCD phase structure from functional methods}},}\ }\href {\doibase 10.1103/PhysRevD.102.034027} {\bibfield  {journal} {\bibinfo  {journal} {Phys. Rev. D}\ }\textbf {\bibinfo {volume} {102}},\ \bibinfo {pages} {034027} (\bibinfo {year} {2020})},\ \Eprint {http://arxiv.org/abs/2002.07500} {arXiv:2002.07500 [hep-ph]} \BibitemShut {NoStop}%
\bibitem [{\citenamefont {jie Fu}(2022)}]{Fu:2022gou}%
  \BibitemOpen
  \bibfield  {author} {\bibinfo {author} {\bibfnamefont {Wei}\ \bibnamefont {jie Fu}},\ }\bibfield  {title} {\enquote {\bibinfo {title} {Qcd at finite temperature and density within the frg approach: an overview},}\ }\href {\doibase 10.1088/1572-9494/ac86be} {\bibfield  {journal} {\bibinfo  {journal} {Commun. Theor. Phys.}\ }\textbf {\bibinfo {volume} {74}},\ \bibinfo {pages} {97304} (\bibinfo {year} {2022})},\ \Eprint {http://arxiv.org/abs/2205.00468} {arXiv:2205.00468 [hep-ph]} \BibitemShut {NoStop}%
\bibitem [{\citenamefont {Kugo}\ and\ \citenamefont {Ojima}(1979)}]{Kugo:1979gm}%
  \BibitemOpen
  \bibfield  {author} {\bibinfo {author} {\bibfnamefont {Taichiro}\ \bibnamefont {Kugo}}\ and\ \bibinfo {author} {\bibfnamefont {Izumi}\ \bibnamefont {Ojima}},\ }\bibfield  {title} {\enquote {\bibinfo {title} {{Local Covariant Operator Formalism of Nonabelian Gauge Theories and Quark Confinement Problem}},}\ }\href@noop {} {\bibfield  {journal} {\bibinfo  {journal} {Prog. Theor. Phys. Suppl.}\ }\textbf {\bibinfo {volume} {66}},\ \bibinfo {pages} {1} (\bibinfo {year} {1979})}\BibitemShut {NoStop}%
%\%CITATION = PTPSA,66,1;\%\%
\bibitem [{\citenamefont {Nakanishi}\ and\ \citenamefont {Ojima}(1990)}]{Nakanishi:1990qm}%
  \BibitemOpen
  \bibfield  {author} {\bibinfo {author} {\bibfnamefont {N.}~\bibnamefont {Nakanishi}}\ and\ \bibinfo {author} {\bibfnamefont {I.}~\bibnamefont {Ojima}},\ }\href@noop {} {\emph {\bibinfo {title} {{Covariant operator formalism of gauge theories and quantum gravity}}}},\ Vol.~\bibinfo {volume} {27}\ (\bibinfo {year} {1990})\BibitemShut {NoStop}%
\bibitem [{\citenamefont {Kugo}(1995)}]{Kugo:1995km}%
  \BibitemOpen
  \bibfield  {author} {\bibinfo {author} {\bibfnamefont {Taichiro}\ \bibnamefont {Kugo}},\ }\bibfield  {title} {\enquote {\bibinfo {title} {{The universal renormalization factors Z(1) / Z(3) and color confinement condition in non-Abelian gauge theory}},}\ }\href {\doibase 10.48550/arxiv.hep-th/9511033} {\  (\bibinfo {year} {1995}),\ 10.48550/arxiv.hep-th/9511033},\ \Eprint {http://arxiv.org/abs/hep-th/9511033} {arXiv:hep-th/9511033} \BibitemShut {NoStop}%
%\%CITATION = HEP-TH/9511033;\%\%
\bibitem [{\citenamefont {von Smekal}\ \emph {et~al.}(1997)\citenamefont {von Smekal}, \citenamefont {Alkofer},\ and\ \citenamefont {Hauck}}]{vonSmekal:1997ohs}%
  \BibitemOpen
  \bibfield  {author} {\bibinfo {author} {\bibfnamefont {Lorenz}\ \bibnamefont {von Smekal}}, \bibinfo {author} {\bibfnamefont {Reinhard}\ \bibnamefont {Alkofer}}, \ and\ \bibinfo {author} {\bibfnamefont {Andreas}\ \bibnamefont {Hauck}},\ }\bibfield  {title} {\enquote {\bibinfo {title} {{The Infrared behavior of gluon and ghost propagators in Landau gauge QCD}},}\ }\href {\doibase 10.1103/PhysRevLett.79.3591} {\bibfield  {journal} {\bibinfo  {journal} {Phys. Rev. Lett.}\ }\textbf {\bibinfo {volume} {79}},\ \bibinfo {pages} {3591--3594} (\bibinfo {year} {1997})},\ \Eprint {http://arxiv.org/abs/hep-ph/9705242} {arXiv:hep-ph/9705242} \BibitemShut {NoStop}%
\bibitem [{\citenamefont {Alkofer}\ and\ \citenamefont {von Smekal}(2001)}]{Alkofer:2000wg}%
  \BibitemOpen
  \bibfield  {author} {\bibinfo {author} {\bibfnamefont {Reinhard}\ \bibnamefont {Alkofer}}\ and\ \bibinfo {author} {\bibfnamefont {Lorenz}\ \bibnamefont {von Smekal}},\ }\bibfield  {title} {\enquote {\bibinfo {title} {{The infrared behavior of QCD Green's functions: Confinement, dynamical symmetry breaking, and hadrons as relativistic bound states}},}\ }\href {\doibase 10.1016/S0370-1573(01)00010-2} {\bibfield  {journal} {\bibinfo  {journal} {Phys. Rept.}\ }\textbf {\bibinfo {volume} {353}},\ \bibinfo {pages} {281} (\bibinfo {year} {2001})},\ \Eprint {http://arxiv.org/abs/hep-ph/0007355} {arXiv:hep-ph/0007355} \BibitemShut {NoStop}%
%\%CITATION = HEP-PH/0007355;\%\%
\bibitem [{\citenamefont {Zwanziger}(2002)}]{Zwanziger:2001kw}%
  \BibitemOpen
  \bibfield  {author} {\bibinfo {author} {\bibfnamefont {Daniel}\ \bibnamefont {Zwanziger}},\ }\bibfield  {title} {\enquote {\bibinfo {title} {{Non-perturbative Landau gauge and infrared critical exponents in QCD}},}\ }\href {\doibase 10.1103/PhysRevD.65.094039} {\bibfield  {journal} {\bibinfo  {journal} {Phys. Rev.}\ }\textbf {\bibinfo {volume} {D65}},\ \bibinfo {pages} {094039} (\bibinfo {year} {2002})},\ \Eprint {http://arxiv.org/abs/hep-th/0109224} {arXiv:hep-th/0109224} \BibitemShut {NoStop}%
%\%CITATION = HEP-TH/0109224;\%\%
\bibitem [{\citenamefont {Lerche}\ and\ \citenamefont {von Smekal}(2002)}]{Lerche:2002ep}%
  \BibitemOpen
  \bibfield  {author} {\bibinfo {author} {\bibfnamefont {Christoph}\ \bibnamefont {Lerche}}\ and\ \bibinfo {author} {\bibfnamefont {Lorenz}\ \bibnamefont {von Smekal}},\ }\bibfield  {title} {\enquote {\bibinfo {title} {{On the infrared exponent for gluon and ghost propagation in Landau gauge QCD}},}\ }\href {\doibase 10.1103/PhysRevD.65.125006} {\bibfield  {journal} {\bibinfo  {journal} {Phys. Rev.}\ }\textbf {\bibinfo {volume} {D65}},\ \bibinfo {pages} {125006} (\bibinfo {year} {2002})},\ \Eprint {http://arxiv.org/abs/hep-ph/0202194} {arXiv:hep-ph/0202194} \BibitemShut {NoStop}%
%\%CITATION = HEP-PH/0202194;\%\%
\bibitem [{\citenamefont {Aguilar}\ \emph {et~al.}(2012)\citenamefont {Aguilar}, \citenamefont {Ibanez}, \citenamefont {Mathieu},\ and\ \citenamefont {Papavassiliou}}]{Aguilar:2011xe}%
  \BibitemOpen
  \bibfield  {author} {\bibinfo {author} {\bibfnamefont {A.~C.}\ \bibnamefont {Aguilar}}, \bibinfo {author} {\bibfnamefont {D.}~\bibnamefont {Ibanez}}, \bibinfo {author} {\bibfnamefont {V.}~\bibnamefont {Mathieu}}, \ and\ \bibinfo {author} {\bibfnamefont {J.}~\bibnamefont {Papavassiliou}},\ }\bibfield  {title} {\enquote {\bibinfo {title} {Massless bound-state excitations and the schwinger mechanism in qcd},}\ }\href {\doibase 10.1103/PhysRevD.85.014018} {\bibfield  {journal} {\bibinfo  {journal} {Phys. Rev. D}\ }\textbf {\bibinfo {volume} {85}},\ \bibinfo {pages} {014018} (\bibinfo {year} {2012})},\ \Eprint {http://arxiv.org/abs/1110.2633} {arXiv:1110.2633 [hep-ph]} \BibitemShut {NoStop}%
\bibitem [{\citenamefont {Aguilar}\ \emph {et~al.}(2022)\citenamefont {Aguilar}, \citenamefont {Ferreira},\ and\ \citenamefont {Papavassiliou}}]{Aguilar:2021uwa}%
  \BibitemOpen
  \bibfield  {author} {\bibinfo {author} {\bibfnamefont {A.~C.}\ \bibnamefont {Aguilar}}, \bibinfo {author} {\bibfnamefont {M.~N.}\ \bibnamefont {Ferreira}}, \ and\ \bibinfo {author} {\bibfnamefont {J.}~\bibnamefont {Papavassiliou}},\ }\bibfield  {title} {\enquote {\bibinfo {title} {Exploring smoking-gun signals of the schwinger mechanism in qcd},}\ }\href {\doibase 10.1103/PhysRevD.105.014030} {\bibfield  {journal} {\bibinfo  {journal} {Phys. Rev. D}\ }\textbf {\bibinfo {volume} {105}},\ \bibinfo {pages} {014030} (\bibinfo {year} {2022})},\ \Eprint {http://arxiv.org/abs/2111.09431} {arXiv:2111.09431 [hep-ph]} \BibitemShut {NoStop}%
\bibitem [{\citenamefont {Aguilar}\ \emph {et~al.}(2023)\citenamefont {Aguilar}, \citenamefont {De~Soto}, \citenamefont {Ferreira}, \citenamefont {Papavassiliou}, \citenamefont {Pinto-G\'omez}, \citenamefont {Roberts},\ and\ \citenamefont {Rodr\'\i{}guez-Quintero}}]{Aguilar:2022thg}%
  \BibitemOpen
  \bibfield  {author} {\bibinfo {author} {\bibfnamefont {A.~C.}\ \bibnamefont {Aguilar}}, \bibinfo {author} {\bibfnamefont {F.}~\bibnamefont {De~Soto}}, \bibinfo {author} {\bibfnamefont {M.~N.}\ \bibnamefont {Ferreira}}, \bibinfo {author} {\bibfnamefont {J.}~\bibnamefont {Papavassiliou}}, \bibinfo {author} {\bibfnamefont {F.}~\bibnamefont {Pinto-G\'omez}}, \bibinfo {author} {\bibfnamefont {C.~D.}\ \bibnamefont {Roberts}}, \ and\ \bibinfo {author} {\bibfnamefont {J.}~\bibnamefont {Rodr\'\i{}guez-Quintero}},\ }\bibfield  {title} {\enquote {\bibinfo {title} {{Schwinger mechanism for gluons from lattice QCD}},}\ }\href {\doibase 10.1016/j.physletb.2023.137906} {\bibfield  {journal} {\bibinfo  {journal} {Phys. Lett. B}\ }\textbf {\bibinfo {volume} {841}},\ \bibinfo {pages} {137906} (\bibinfo {year} {2023})},\ \Eprint {http://arxiv.org/abs/2211.12594} {arXiv:2211.12594 [hep-ph]} \BibitemShut {NoStop}%
\bibitem [{\citenamefont {Ferreira}\ and\ \citenamefont {Papavassiliou}(2025)}]{Ferreira:2025anh}%
  \BibitemOpen
  \bibfield  {author} {\bibinfo {author} {\bibfnamefont {M.~N.}\ \bibnamefont {Ferreira}}\ and\ \bibinfo {author} {\bibfnamefont {J.}~\bibnamefont {Papavassiliou}},\ }\bibfield  {title} {\enquote {\bibinfo {title} {Gluon mass scale through the schwinger mechanism},}\ }\href@noop {} {\  (\bibinfo {year} {2025})},\ \Eprint {http://arxiv.org/abs/2501.01080} {arXiv:2501.01080 [hep-ph]} \BibitemShut {NoStop}%
\bibitem [{\citenamefont {Alkofer}\ and\ \citenamefont {Alkofer}(2011)}]{Alkofer:2011pe}%
  \BibitemOpen
  \bibfield  {author} {\bibinfo {author} {\bibfnamefont {Natalia}\ \bibnamefont {Alkofer}}\ and\ \bibinfo {author} {\bibfnamefont {Reinhard}\ \bibnamefont {Alkofer}},\ }\bibfield  {title} {\enquote {\bibinfo {title} {{Features of ghost-gluon and ghost-quark bound states related to BRST quartets}},}\ }\href {\doibase 10.48550/arxiv.1102.2753} {\  (\bibinfo {year} {2011}),\ 10.48550/arxiv.1102.2753},\ \bibinfo {note} {kugo-Ojima quartet mechanism},\ \Eprint {http://arxiv.org/abs/1102.2753} {arXiv:1102.2753 [hep-th]} \BibitemShut {NoStop}%
%\%CITATION = 1102.2753;\%\%
\bibitem [{\citenamefont {Fu}\ \emph {et~al.}(2025)\citenamefont {Fu}, \citenamefont {Huang}, \citenamefont {Pawlowski}, \citenamefont {Tan},\ and\ \citenamefont {Zhou}}]{Fu:2025hcm}%
  \BibitemOpen
  \bibfield  {author} {\bibinfo {author} {\bibfnamefont {Wei-jie}\ \bibnamefont {Fu}}, \bibinfo {author} {\bibfnamefont {Chuang}\ \bibnamefont {Huang}}, \bibinfo {author} {\bibfnamefont {Jan~M.}\ \bibnamefont {Pawlowski}}, \bibinfo {author} {\bibfnamefont {Yang-yang}\ \bibnamefont {Tan}}, \ and\ \bibinfo {author} {\bibfnamefont {Li-jun}\ \bibnamefont {Zhou}},\ }\bibfield  {title} {\enquote {\bibinfo {title} {{Four-quark scatterings in QCD III}},}\ }\href@noop {} {\  (\bibinfo {year} {2025})},\ \Eprint {http://arxiv.org/abs/2502.14388} {arXiv:2502.14388 [hep-ph]} \BibitemShut {NoStop}%
\bibitem [{\citenamefont {Fischer}\ \emph {et~al.}(2009)\citenamefont {Fischer}, \citenamefont {Maas},\ and\ \citenamefont {Pawlowski}}]{Fischer:2008uz}%
  \BibitemOpen
  \bibfield  {author} {\bibinfo {author} {\bibfnamefont {Christian~S.}\ \bibnamefont {Fischer}}, \bibinfo {author} {\bibfnamefont {Axel}\ \bibnamefont {Maas}}, \ and\ \bibinfo {author} {\bibfnamefont {Jan~M.}\ \bibnamefont {Pawlowski}},\ }\bibfield  {title} {\enquote {\bibinfo {title} {{On the infrared behavior of Landau gauge Yang-Mills theory}},}\ }\href {\doibase 10.1016/j.aop.2009.07.009} {\bibfield  {journal} {\bibinfo  {journal} {Annals Phys.}\ }\textbf {\bibinfo {volume} {324}},\ \bibinfo {pages} {2408--2437} (\bibinfo {year} {2009})},\ \Eprint {http://arxiv.org/abs/0810.1987} {arXiv:0810.1987 [hep-ph]} \BibitemShut {NoStop}%
%\%CITATION = 0810.1987;\%\%
\bibitem [{\citenamefont {Pawlowski}\ \emph {et~al.}(2004)\citenamefont {Pawlowski}, \citenamefont {Litim}, \citenamefont {Nedelko},\ and\ \citenamefont {von Smekal}}]{Pawlowski:2003hq}%
  \BibitemOpen
  \bibfield  {author} {\bibinfo {author} {\bibfnamefont {Jan~M.}\ \bibnamefont {Pawlowski}}, \bibinfo {author} {\bibfnamefont {Daniel~F.}\ \bibnamefont {Litim}}, \bibinfo {author} {\bibfnamefont {Sergei}\ \bibnamefont {Nedelko}}, \ and\ \bibinfo {author} {\bibfnamefont {Lorenz}\ \bibnamefont {von Smekal}},\ }\bibfield  {title} {\enquote {\bibinfo {title} {{Infrared behavior and fixed points in Landau gauge QCD}},}\ }\href {\doibase 10.1103/PhysRevLett.93.152002} {\bibfield  {journal} {\bibinfo  {journal} {Phys.Rev.Lett.}\ }\textbf {\bibinfo {volume} {93}},\ \bibinfo {pages} {152002} (\bibinfo {year} {2004})},\ \Eprint {http://arxiv.org/abs/hep-th/0312324} {arXiv:hep-th/0312324 [hep-th]} \BibitemShut {NoStop}%
%\%CITATION = HEP-TH/0312324;\%\%
\bibitem [{\citenamefont {Ferreira}\ \emph {et~al.}(2025)\citenamefont {Ferreira}, \citenamefont {Papavassiliou}, \citenamefont {Pawlowski},\ and\ \citenamefont {Wink}}]{Ferreira:2025tzo}%
  \BibitemOpen
  \bibfield  {author} {\bibinfo {author} {\bibfnamefont {Mauricio~N.}\ \bibnamefont {Ferreira}}, \bibinfo {author} {\bibfnamefont {Joannis}\ \bibnamefont {Papavassiliou}}, \bibinfo {author} {\bibfnamefont {Jan~M.}\ \bibnamefont {Pawlowski}}, \ and\ \bibinfo {author} {\bibfnamefont {Nicolas}\ \bibnamefont {Wink}},\ }\bibfield  {title} {\enquote {\bibinfo {title} {{Physics of the gluon mass gap}},}\ }\href {\doibase 10.1140/epjc/s10052-025-15027-7} {\bibfield  {journal} {\bibinfo  {journal} {Eur. Phys. J. C}\ }\textbf {\bibinfo {volume} {85}},\ \bibinfo {pages} {1339} (\bibinfo {year} {2025})},\ \Eprint {http://arxiv.org/abs/2508.20568} {arXiv:2508.20568 [hep-ph]} \BibitemShut {NoStop}%
\bibitem [{\citenamefont {Nambu}\ and\ \citenamefont {Jona-Lasinio}(1961)}]{Nambu:1961fr}%
  \BibitemOpen
  \bibfield  {author} {\bibinfo {author} {\bibfnamefont {Yoichiro}\ \bibnamefont {Nambu}}\ and\ \bibinfo {author} {\bibfnamefont {G.}~\bibnamefont {Jona-Lasinio}},\ }\bibfield  {title} {\enquote {\bibinfo {title} {{Dynamical model of elementary particles based on an analogy with superconductivity. II.}}}\ }\href {\doibase 10.1103/PhysRev.124.246} {\bibfield  {journal} {\bibinfo  {journal} {Phys. Rev.}\ }\textbf {\bibinfo {volume} {124}},\ \bibinfo {pages} {246--254} (\bibinfo {year} {1961})}\BibitemShut {NoStop}%
\bibitem [{\citenamefont {Hubbard}(1959)}]{HubbardPhysRevLett.3.77}%
  \BibitemOpen
  \bibfield  {author} {\bibinfo {author} {\bibfnamefont {J.}~\bibnamefont {Hubbard}},\ }\bibfield  {title} {\enquote {\bibinfo {title} {Calculation of partition functions},}\ }\href {\doibase 10.1103/PhysRevLett.3.77} {\bibfield  {journal} {\bibinfo  {journal} {Phys. Rev. Lett.}\ }\textbf {\bibinfo {volume} {3}},\ \bibinfo {pages} {77--78} (\bibinfo {year} {1959})}\BibitemShut {NoStop}%
\bibitem [{\citenamefont {{Stratonovich}}(1957)}]{Stratonovich}%
  \BibitemOpen
  \bibfield  {author} {\bibinfo {author} {\bibfnamefont {R.~L.}\ \bibnamefont {{Stratonovich}}},\ }\bibfield  {title} {\enquote {\bibinfo {title} {{On a Method of Calculating Quantum Distribution Functions}},}\ }\href@noop {} {\bibfield  {journal} {\bibinfo  {journal} {Soviet Physics Doklady}\ }\textbf {\bibinfo {volume} {2}},\ \bibinfo {pages} {416} (\bibinfo {year} {1957})}\BibitemShut {NoStop}%
\bibitem [{\citenamefont {Miransky}\ and\ \citenamefont {Yamawaki}(1989)}]{Miransky1989b}%
  \BibitemOpen
  \bibfield  {author} {\bibinfo {author} {\bibfnamefont {V.~A.}\ \bibnamefont {Miransky}}\ and\ \bibinfo {author} {\bibfnamefont {K.}~\bibnamefont {Yamawaki}},\ }\bibfield  {title} {\enquote {\bibinfo {title} {{On Gauge Theories with Additional Four Fermion Interaction}},}\ }\href {\doibase 10.1142/S0217732389000186} {\bibfield  {journal} {\bibinfo  {journal} {Mod. Phys. Lett. A}\ }\textbf {\bibinfo {volume} {4}},\ \bibinfo {pages} {129--135} (\bibinfo {year} {1989})}\BibitemShut {NoStop}%
\bibitem [{\citenamefont {Miransky}(1994)}]{Miransky:1994vk}%
  \BibitemOpen
  \bibfield  {author} {\bibinfo {author} {\bibfnamefont {V.~A.}\ \bibnamefont {Miransky}},\ }\href@noop {} {\emph {\bibinfo {title} {{Dynamical symmetry breaking in quantum field theories}}}}\ (\bibinfo {year} {1994})\BibitemShut {NoStop}%
\bibitem [{\citenamefont {Gies}\ and\ \citenamefont {Jaeckel}(2006)}]{Gies:2005as}%
  \BibitemOpen
  \bibfield  {author} {\bibinfo {author} {\bibfnamefont {Holger}\ \bibnamefont {Gies}}\ and\ \bibinfo {author} {\bibfnamefont {Joerg}\ \bibnamefont {Jaeckel}},\ }\bibfield  {title} {\enquote {\bibinfo {title} {{Chiral phase structure of QCD with many flavors}},}\ }\href {\doibase 10.1140/epjc/s2006-02475-0} {\bibfield  {journal} {\bibinfo  {journal} {Eur. Phys. J. C}\ }\textbf {\bibinfo {volume} {46}},\ \bibinfo {pages} {433--438} (\bibinfo {year} {2006})},\ \Eprint {http://arxiv.org/abs/hep-ph/0507171} {arXiv:hep-ph/0507171} \BibitemShut {NoStop}%
\bibitem [{\citenamefont {Braun}(2012)}]{Braun:2011pp}%
  \BibitemOpen
  \bibfield  {author} {\bibinfo {author} {\bibfnamefont {Jens}\ \bibnamefont {Braun}},\ }\bibfield  {title} {\enquote {\bibinfo {title} {{Fermion Interactions and Universal Behavior in Strongly Interacting Theories}},}\ }\href {\doibase 10.1088/0954-3899/39/3/033001} {\bibfield  {journal} {\bibinfo  {journal} {J.Phys.}\ }\textbf {\bibinfo {volume} {G39}},\ \bibinfo {pages} {033001} (\bibinfo {year} {2012})},\ \Eprint {http://arxiv.org/abs/1108.4449} {arXiv:1108.4449 [hep-ph]} \BibitemShut {NoStop}%
%\%CITATION = ARXIV:1108.4449;\%\%
\bibitem [{\citenamefont {Gies}\ and\ \citenamefont {Wetterich}(2004)}]{Gies:2002hq}%
  \BibitemOpen
  \bibfield  {author} {\bibinfo {author} {\bibfnamefont {Holger}\ \bibnamefont {Gies}}\ and\ \bibinfo {author} {\bibfnamefont {Christof}\ \bibnamefont {Wetterich}},\ }\bibfield  {title} {\enquote {\bibinfo {title} {{Universality of spontaneous chiral symmetry breaking in gauge theories}},}\ }\href {\doibase 10.1103/PhysRevD.69.025001} {\bibfield  {journal} {\bibinfo  {journal} {Phys.Rev.}\ }\textbf {\bibinfo {volume} {D69}},\ \bibinfo {pages} {025001} (\bibinfo {year} {2004})},\ \Eprint {http://arxiv.org/abs/hep-th/0209183} {arXiv:hep-th/0209183 [hep-th]} \BibitemShut {NoStop}%
%\%CITATION = HEP-TH/0209183;\%\%
\bibitem [{\citenamefont {Pawlowski}(2007)}]{Pawlowski:2005xe}%
  \BibitemOpen
  \bibfield  {author} {\bibinfo {author} {\bibfnamefont {Jan~M.}\ \bibnamefont {Pawlowski}},\ }\bibfield  {title} {\enquote {\bibinfo {title} {{Aspects of the functional renormalisation group}},}\ }\href {\doibase 10.1016/j.aop.2007.01.007} {\bibfield  {journal} {\bibinfo  {journal} {Annals Phys.}\ }\textbf {\bibinfo {volume} {322}},\ \bibinfo {pages} {2831--2915} (\bibinfo {year} {2007})},\ \Eprint {http://arxiv.org/abs/hep-th/0512261} {arXiv:hep-th/0512261 [hep-th]} \BibitemShut {NoStop}%
%\%CITATION = HEP-TH/0512261;\%\%
\bibitem [{\citenamefont {Braun}\ \emph {et~al.}(2016)\citenamefont {Braun}, \citenamefont {Fister}, \citenamefont {Pawlowski},\ and\ \citenamefont {Rennecke}}]{Braun:2014ata}%
  \BibitemOpen
  \bibfield  {author} {\bibinfo {author} {\bibfnamefont {Jens}\ \bibnamefont {Braun}}, \bibinfo {author} {\bibfnamefont {Leonard}\ \bibnamefont {Fister}}, \bibinfo {author} {\bibfnamefont {Jan~M}\ \bibnamefont {Pawlowski}}, \ and\ \bibinfo {author} {\bibfnamefont {Fabian}\ \bibnamefont {Rennecke}},\ }\bibfield  {title} {\enquote {\bibinfo {title} {From quarks and gluons to hadrons: Chiral symmetry breaking in dynamical qcd},}\ }\href {\doibase 10.1103/PhysRevD.94.034016} {\bibfield  {journal} {\bibinfo  {journal} {Phys. Rev. D}\ }\textbf {\bibinfo {volume} {94}},\ \bibinfo {pages} {34016} (\bibinfo {year} {2016})},\ \Eprint {http://arxiv.org/abs/1412.1045} {arXiv:1412.1045 [hep-ph]} \BibitemShut {NoStop}%
\bibitem [{\citenamefont {Huber}(2020{\natexlab{a}})}]{Huber:2020keu}%
  \BibitemOpen
  \bibfield  {author} {\bibinfo {author} {\bibfnamefont {Markus~Q.}\ \bibnamefont {Huber}},\ }\bibfield  {title} {\enquote {\bibinfo {title} {{Correlation functions of Landau gauge Yang-Mills theory}},}\ }\href {\doibase 10.1103/PhysRevD.101.114009} {\bibfield  {journal} {\bibinfo  {journal} {Phys. Rev. D}\ }\textbf {\bibinfo {volume} {101}},\ \bibinfo {pages} {114009} (\bibinfo {year} {2020}{\natexlab{a}})},\ \Eprint {http://arxiv.org/abs/2003.13703} {arXiv:2003.13703 [hep-ph]} \BibitemShut {NoStop}%
\bibitem [{\citenamefont {Huber}(2020{\natexlab{b}})}]{Huber:2018ned}%
  \BibitemOpen
  \bibfield  {author} {\bibinfo {author} {\bibfnamefont {Markus~Q.}\ \bibnamefont {Huber}},\ }\bibfield  {title} {\enquote {\bibinfo {title} {{Nonperturbative properties of Yang{\textendash}Mills theories}},}\ }\href {\doibase 10.1016/j.physrep.2020.04.004} {\bibfield  {journal} {\bibinfo  {journal} {Phys. Rept.}\ }\textbf {\bibinfo {volume} {879}},\ \bibinfo {pages} {1--92} (\bibinfo {year} {2020}{\natexlab{b}})},\ \Eprint {http://arxiv.org/abs/1808.05227} {arXiv:1808.05227 [hep-ph]} \BibitemShut {NoStop}%
\bibitem [{\citenamefont {Fischer}\ and\ \citenamefont {Alkofer}(2002)}]{Fischer:2002hna}%
  \BibitemOpen
  \bibfield  {author} {\bibinfo {author} {\bibfnamefont {C.~S.}\ \bibnamefont {Fischer}}\ and\ \bibinfo {author} {\bibfnamefont {Reinhard}\ \bibnamefont {Alkofer}},\ }\bibfield  {title} {\enquote {\bibinfo {title} {{Infrared exponents and running coupling of SU(N) Yang- Mills theories}},}\ }\href {\doibase 10.1016/S0370-2693(02)01809-9} {\bibfield  {journal} {\bibinfo  {journal} {Phys. Lett.}\ }\textbf {\bibinfo {volume} {B536}},\ \bibinfo {pages} {177--184} (\bibinfo {year} {2002})},\ \Eprint {http://arxiv.org/abs/hep-ph/0202202} {arXiv:hep-ph/0202202} \BibitemShut {NoStop}%
%\%CITATION = HEP-PH/0202202;\%\%
\bibitem [{\citenamefont {Eichmann}\ \emph {et~al.}(2014)\citenamefont {Eichmann}, \citenamefont {Williams}, \citenamefont {Alkofer},\ and\ \citenamefont {Vujinovic}}]{Eichmann:2014xya}%
  \BibitemOpen
  \bibfield  {author} {\bibinfo {author} {\bibfnamefont {Gernot}\ \bibnamefont {Eichmann}}, \bibinfo {author} {\bibfnamefont {Richard}\ \bibnamefont {Williams}}, \bibinfo {author} {\bibfnamefont {Reinhard}\ \bibnamefont {Alkofer}}, \ and\ \bibinfo {author} {\bibfnamefont {Milan}\ \bibnamefont {Vujinovic}},\ }\bibfield  {title} {\enquote {\bibinfo {title} {{The three-gluon vertex in Landau gauge}},}\ }\href {\doibase 10.1103/PhysRevD.89.105014} {\bibfield  {journal} {\bibinfo  {journal} {Phys.Rev.}\ }\textbf {\bibinfo {volume} {D89}},\ \bibinfo {pages} {105014} (\bibinfo {year} {2014})},\ \Eprint {http://arxiv.org/abs/1402.1365} {arXiv:1402.1365 [hep-ph]} \BibitemShut {NoStop}%
%%CITATION = ARXIV:1402.1365;%%
\bibitem [{\citenamefont {Papavassiliou}\ \emph {et~al.}(2022)\citenamefont {Papavassiliou}, \citenamefont {Aguilar},\ and\ \citenamefont {Ferreira}}]{Papavassiliou:2022umz}%
  \BibitemOpen
  \bibfield  {author} {\bibinfo {author} {\bibfnamefont {Joannis}\ \bibnamefont {Papavassiliou}}, \bibinfo {author} {\bibfnamefont {A.~C.}\ \bibnamefont {Aguilar}}, \ and\ \bibinfo {author} {\bibfnamefont {M.~N.}\ \bibnamefont {Ferreira}},\ }\bibfield  {title} {\enquote {\bibinfo {title} {{Theory and phenomenology of the three-gluon vertex}},}\ }\href {\doibase 10.31349/SuplRevMexFis.3.0308112} {\bibfield  {journal} {\bibinfo  {journal} {Rev. Mex. Fis. Suppl.}\ }\textbf {\bibinfo {volume} {3}},\ \bibinfo {pages} {0308112} (\bibinfo {year} {2022})},\ \Eprint {http://arxiv.org/abs/2201.08496} {arXiv:2201.08496 [hep-ph]} \BibitemShut {NoStop}%
\bibitem [{\citenamefont {Athenodorou}\ \emph {et~al.}(2016)\citenamefont {Athenodorou}, \citenamefont {Binosi}, \citenamefont {Boucaud}, \citenamefont {De~Soto}, \citenamefont {Papavassiliou}, \citenamefont {Rodriguez-Quintero},\ and\ \citenamefont {Zafeiropoulos}}]{Athenodorou:2016oyh}%
  \BibitemOpen
  \bibfield  {author} {\bibinfo {author} {\bibfnamefont {A.}~\bibnamefont {Athenodorou}}, \bibinfo {author} {\bibfnamefont {D.}~\bibnamefont {Binosi}}, \bibinfo {author} {\bibfnamefont {Ph.}\ \bibnamefont {Boucaud}}, \bibinfo {author} {\bibfnamefont {F.}~\bibnamefont {De~Soto}}, \bibinfo {author} {\bibfnamefont {J.}~\bibnamefont {Papavassiliou}}, \bibinfo {author} {\bibfnamefont {J.}~\bibnamefont {Rodriguez-Quintero}}, \ and\ \bibinfo {author} {\bibfnamefont {S.}~\bibnamefont {Zafeiropoulos}},\ }\bibfield  {title} {\enquote {\bibinfo {title} {{On the zero crossing of the three-gluon vertex}},}\ }\href {\doibase 10.1016/j.physletb.2016.08.065} {\bibfield  {journal} {\bibinfo  {journal} {Phys. Lett. B}\ }\textbf {\bibinfo {volume} {761}},\ \bibinfo {pages} {444--449} (\bibinfo {year} {2016})},\ \Eprint {http://arxiv.org/abs/1607.01278} {arXiv:1607.01278 [hep-ph]} \BibitemShut {NoStop}%
\bibitem [{\citenamefont {Pinto-G{\'o}mez}\ \emph {et~al.}(2024)\citenamefont {Pinto-G{\'o}mez}, \citenamefont {De~Soto},\ and\ \citenamefont {Rodr{\'\i}guez-Quintero}}]{Pinto-Gomez:2024mrk}%
  \BibitemOpen
  \bibfield  {author} {\bibinfo {author} {\bibfnamefont {F.}~\bibnamefont {Pinto-G{\'o}mez}}, \bibinfo {author} {\bibfnamefont {F.}~\bibnamefont {De~Soto}}, \ and\ \bibinfo {author} {\bibfnamefont {J.}~\bibnamefont {Rodr{\'\i}guez-Quintero}},\ }\bibfield  {title} {\enquote {\bibinfo {title} {{Complete analysis of the Landau-gauge three-gluon vertex from lattice QCD}},}\ }\href {\doibase 10.1103/PhysRevD.110.014005} {\bibfield  {journal} {\bibinfo  {journal} {Phys. Rev. D}\ }\textbf {\bibinfo {volume} {110}},\ \bibinfo {pages} {014005} (\bibinfo {year} {2024})},\ \Eprint {http://arxiv.org/abs/2404.08777} {arXiv:2404.08777 [hep-ph]} \BibitemShut {NoStop}%
\bibitem [{\citenamefont {Brito}\ \emph {et~al.}(2025)\citenamefont {Brito}, \citenamefont {Oliveira},\ and\ \citenamefont {Silva}}]{Brito:2024aod}%
  \BibitemOpen
  \bibfield  {author} {\bibinfo {author} {\bibfnamefont {Nuno}\ \bibnamefont {Brito}}, \bibinfo {author} {\bibfnamefont {Orlando}\ \bibnamefont {Oliveira}}, \ and\ \bibinfo {author} {\bibfnamefont {Paulo~J.}\ \bibnamefont {Silva}},\ }\bibfield  {title} {\enquote {\bibinfo {title} {{High statistical computation of the Landau gauge ghost-gluon vertex}},}\ }\href {\doibase 10.22323/1.466.0469} {\bibfield  {journal} {\bibinfo  {journal} {PoS}\ }\textbf {\bibinfo {volume} {LATTICE2024}},\ \bibinfo {pages} {469} (\bibinfo {year} {2025})},\ \Eprint {http://arxiv.org/abs/2411.17280} {arXiv:2411.17280 [hep-lat]} \BibitemShut {NoStop}%
\bibitem [{\citenamefont {Ilgenfritz}\ \emph {et~al.}(2007)\citenamefont {Ilgenfritz}, \citenamefont {M{\"u}ller-Preussker}, \citenamefont {Sternbeck}, \citenamefont {Schiller},\ and\ \citenamefont {Bogolubsky}}]{Ilgenfritz:2006he}%
  \BibitemOpen
  \bibfield  {author} {\bibinfo {author} {\bibfnamefont {E.~M.}\ \bibnamefont {Ilgenfritz}}, \bibinfo {author} {\bibfnamefont {M.}~\bibnamefont {M{\"u}ller-Preussker}}, \bibinfo {author} {\bibfnamefont {A.}~\bibnamefont {Sternbeck}}, \bibinfo {author} {\bibfnamefont {A.}~\bibnamefont {Schiller}}, \ and\ \bibinfo {author} {\bibfnamefont {I.~L.}\ \bibnamefont {Bogolubsky}},\ }\bibfield  {title} {\enquote {\bibinfo {title} {{Landau gauge gluon and ghost propagators from lattice QCD}},}\ }\href {\doibase 10.48550/arxiv.hep-lat/0609043} {\bibfield  {journal} {\bibinfo  {journal} {Braz. J. Phys.}\ }\textbf {\bibinfo {volume} {37}},\ \bibinfo {pages} {193} (\bibinfo {year} {2007})},\ \Eprint {http://arxiv.org/abs/hep-lat/0609043} {arXiv:hep-lat/0609043} \BibitemShut {NoStop}%
%\%CITATION = HEP-LAT/0609043;\%\%
\bibitem [{\citenamefont {Dimopoulos}\ \emph {et~al.}(1980)\citenamefont {Dimopoulos}, \citenamefont {Raby},\ and\ \citenamefont {Susskind}}]{Dimopoulos:1980hn}%
  \BibitemOpen
  \bibfield  {author} {\bibinfo {author} {\bibfnamefont {S.}~\bibnamefont {Dimopoulos}}, \bibinfo {author} {\bibfnamefont {S.}~\bibnamefont {Raby}}, \ and\ \bibinfo {author} {\bibfnamefont {Leonard}\ \bibnamefont {Susskind}},\ }\bibfield  {title} {\enquote {\bibinfo {title} {{Light Composite Fermions}},}\ }\href {\doibase 10.1016/0550-3213(80)90215-1} {\bibfield  {journal} {\bibinfo  {journal} {Nucl. Phys. B}\ }\textbf {\bibinfo {volume} {173}},\ \bibinfo {pages} {208--228} (\bibinfo {year} {1980})}\BibitemShut {NoStop}%
\bibitem [{\citenamefont {Eichten}\ \emph {et~al.}(1986)\citenamefont {Eichten}, \citenamefont {Peccei}, \citenamefont {Preskill},\ and\ \citenamefont {Zeppenfeld}}]{Eichten:1985fs}%
  \BibitemOpen
  \bibfield  {author} {\bibinfo {author} {\bibfnamefont {Estia}\ \bibnamefont {Eichten}}, \bibinfo {author} {\bibfnamefont {Roberto~D.}\ \bibnamefont {Peccei}}, \bibinfo {author} {\bibfnamefont {John}\ \bibnamefont {Preskill}}, \ and\ \bibinfo {author} {\bibfnamefont {Dieter}\ \bibnamefont {Zeppenfeld}},\ }\bibfield  {title} {\enquote {\bibinfo {title} {{Chiral Gauge Theories in the 1/n Expansion}},}\ }\href {\doibase 10.1016/0550-3213(86)90206-3} {\bibfield  {journal} {\bibinfo  {journal} {Nucl. Phys. B}\ }\textbf {\bibinfo {volume} {268}},\ \bibinfo {pages} {161--178} (\bibinfo {year} {1986})}\BibitemShut {NoStop}%
\bibitem [{\citenamefont {Karasik}\ \emph {et~al.}(2022)\citenamefont {Karasik}, \citenamefont {{\"O}nder},\ and\ \citenamefont {Tong}}]{Karasik:2022gve}%
  \BibitemOpen
  \bibfield  {author} {\bibinfo {author} {\bibfnamefont {Avner}\ \bibnamefont {Karasik}}, \bibinfo {author} {\bibfnamefont {Kaan}\ \bibnamefont {{\"O}nder}}, \ and\ \bibinfo {author} {\bibfnamefont {David}\ \bibnamefont {Tong}},\ }\bibfield  {title} {\enquote {\bibinfo {title} {{Chiral gauge dynamics: candidates for non-supersymmetric dualities}},}\ }\href {\doibase 10.1007/JHEP11(2022)122} {\bibfield  {journal} {\bibinfo  {journal} {JHEP}\ }\textbf {\bibinfo {volume} {11}},\ \bibinfo {pages} {122} (\bibinfo {year} {2022})},\ \Eprint {http://arxiv.org/abs/2208.07842} {arXiv:2208.07842 [hep-th]} \BibitemShut {NoStop}%
\bibitem [{\citenamefont {Napetschnig}\ \emph {et~al.}(2021)\citenamefont {Napetschnig}, \citenamefont {Alkofer}, \citenamefont {Huber},\ and\ \citenamefont {Pawlowski}}]{Napetschnig:2021ria}%
  \BibitemOpen
  \bibfield  {author} {\bibinfo {author} {\bibfnamefont {Martin}\ \bibnamefont {Napetschnig}}, \bibinfo {author} {\bibfnamefont {Reinhard}\ \bibnamefont {Alkofer}}, \bibinfo {author} {\bibfnamefont {Markus~Q.}\ \bibnamefont {Huber}}, \ and\ \bibinfo {author} {\bibfnamefont {Jan~M.}\ \bibnamefont {Pawlowski}},\ }\bibfield  {title} {\enquote {\bibinfo {title} {{Yang-Mills propagators in linear covariant gauges from Nielsen identities}},}\ }\href {\doibase 10.1103/PhysRevD.104.054003} {\bibfield  {journal} {\bibinfo  {journal} {Phys. Rev. D}\ }\textbf {\bibinfo {volume} {104}},\ \bibinfo {pages} {054003} (\bibinfo {year} {2021})},\ \Eprint {http://arxiv.org/abs/2106.12559} {arXiv:2106.12559 [hep-ph]} \BibitemShut {NoStop}%
\bibitem [{\citenamefont {Huber}(2015)}]{Huber:2015ria}%
  \BibitemOpen
  \bibfield  {author} {\bibinfo {author} {\bibfnamefont {Markus~Q.}\ \bibnamefont {Huber}},\ }\bibfield  {title} {\enquote {\bibinfo {title} {{Gluon and ghost propagators in linear covariant gauges}},}\ }\href {\doibase 10.1103/PhysRevD.91.085018} {\bibfield  {journal} {\bibinfo  {journal} {Phys. Rev. D}\ }\textbf {\bibinfo {volume} {91}},\ \bibinfo {pages} {085018} (\bibinfo {year} {2015})},\ \Eprint {http://arxiv.org/abs/1502.04057} {arXiv:1502.04057 [hep-ph]} \BibitemShut {NoStop}%
\bibitem [{\citenamefont {Aguilar}\ \emph {et~al.}(2015)\citenamefont {Aguilar}, \citenamefont {Binosi},\ and\ \citenamefont {Papavassiliou}}]{Aguilar:2015nqa}%
  \BibitemOpen
  \bibfield  {author} {\bibinfo {author} {\bibfnamefont {A.~C.}\ \bibnamefont {Aguilar}}, \bibinfo {author} {\bibfnamefont {D.}~\bibnamefont {Binosi}}, \ and\ \bibinfo {author} {\bibfnamefont {J.}~\bibnamefont {Papavassiliou}},\ }\bibfield  {title} {\enquote {\bibinfo {title} {{Yang-Mills two-point functions in linear covariant gauges}},}\ }\href {\doibase 10.1103/PhysRevD.91.085014} {\bibfield  {journal} {\bibinfo  {journal} {Phys. Rev. D}\ }\textbf {\bibinfo {volume} {91}},\ \bibinfo {pages} {085014} (\bibinfo {year} {2015})},\ \Eprint {http://arxiv.org/abs/1501.07150} {arXiv:1501.07150 [hep-ph]} \BibitemShut {NoStop}%
\bibitem [{\citenamefont {Alkofer}\ \emph {et~al.}(2003)\citenamefont {Alkofer}, \citenamefont {Fischer}, \citenamefont {Reinhardt},\ and\ \citenamefont {von Smekal}}]{Alkofer:2003jr}%
  \BibitemOpen
  \bibfield  {author} {\bibinfo {author} {\bibfnamefont {Reinhard}\ \bibnamefont {Alkofer}}, \bibinfo {author} {\bibfnamefont {C.~S.}\ \bibnamefont {Fischer}}, \bibinfo {author} {\bibfnamefont {H.}~\bibnamefont {Reinhardt}}, \ and\ \bibinfo {author} {\bibfnamefont {L.}~\bibnamefont {von Smekal}},\ }\bibfield  {title} {\enquote {\bibinfo {title} {{On the infrared behaviour of gluons and ghosts in ghost- antighost symmetric gauges}},}\ }\href {\doibase 10.1103/PhysRevD.68.045003} {\bibfield  {journal} {\bibinfo  {journal} {Phys. Rev.}\ }\textbf {\bibinfo {volume} {D68}},\ \bibinfo {pages} {045003} (\bibinfo {year} {2003})},\ \Eprint {http://arxiv.org/abs/hep-th/0304134} {arXiv:hep-th/0304134} \BibitemShut {NoStop}%
%\%CITATION = HEP-TH/0304134;\%\%
\bibitem [{\citenamefont {Litim}(2001)}]{Litim:2001up}%
  \BibitemOpen
  \bibfield  {author} {\bibinfo {author} {\bibfnamefont {Daniel~F.}\ \bibnamefont {Litim}},\ }\bibfield  {title} {\enquote {\bibinfo {title} {{Optimized renormalization group flows}},}\ }\href {\doibase 10.1103/PhysRevD.64.105007} {\bibfield  {journal} {\bibinfo  {journal} {Phys.Rev.}\ }\textbf {\bibinfo {volume} {D64}},\ \bibinfo {pages} {105007} (\bibinfo {year} {2001})},\ \Eprint {http://arxiv.org/abs/hep-th/0103195} {arXiv:hep-th/0103195 [hep-th]} \BibitemShut {NoStop}%
%\%CITATION = HEP-TH/0103195;\%\%
\bibitem [{\citenamefont {Ferreira}\ and\ \citenamefont {Papavassiliou}(2023)}]{Ferreira:2023fva}%
  \BibitemOpen
  \bibfield  {author} {\bibinfo {author} {\bibfnamefont {Mauricio~Narciso}\ \bibnamefont {Ferreira}}\ and\ \bibinfo {author} {\bibfnamefont {Joannis}\ \bibnamefont {Papavassiliou}},\ }\bibfield  {title} {\enquote {\bibinfo {title} {Gauge sector dynamics in qcd},}\ }\href {\doibase 10.3390/particles6010017} {\bibfield  {journal} {\bibinfo  {journal} {Particles}\ }\textbf {\bibinfo {volume} {6}},\ \bibinfo {pages} {312--363} (\bibinfo {year} {2023})},\ \Eprint {http://arxiv.org/abs/2301.02314} {arXiv:2301.02314 [hep-ph]} \BibitemShut {NoStop}%
\bibitem [{\citenamefont {Aguilar}\ \emph {et~al.}(2008)\citenamefont {Aguilar}, \citenamefont {Binosi},\ and\ \citenamefont {Papavassiliou}}]{Aguilar:2008xm}%
  \BibitemOpen
  \bibfield  {author} {\bibinfo {author} {\bibfnamefont {A.~C.}\ \bibnamefont {Aguilar}}, \bibinfo {author} {\bibfnamefont {D.}~\bibnamefont {Binosi}}, \ and\ \bibinfo {author} {\bibfnamefont {J.}~\bibnamefont {Papavassiliou}},\ }\bibfield  {title} {\enquote {\bibinfo {title} {{Gluon and ghost propagators in the Landau gauge: Deriving lattice results from Schwinger-Dyson equations}},}\ }\href {\doibase 10.1103/PhysRevD.78.025010} {\bibfield  {journal} {\bibinfo  {journal} {Phys. Rev.}\ }\textbf {\bibinfo {volume} {D78}},\ \bibinfo {pages} {025010} (\bibinfo {year} {2008})},\ \Eprint {http://arxiv.org/abs/0802.1870} {arXiv:0802.1870 [hep-ph]} \BibitemShut {NoStop}%
%\%CITATION = 0802.1870;\%\%
\bibitem [{\citenamefont {Aguilar}\ and\ \citenamefont {Natale}(2004)}]{Aguilar:2004sw}%
  \BibitemOpen
  \bibfield  {author} {\bibinfo {author} {\bibfnamefont {A.~C.}\ \bibnamefont {Aguilar}}\ and\ \bibinfo {author} {\bibfnamefont {A.~A.}\ \bibnamefont {Natale}},\ }\bibfield  {title} {\enquote {\bibinfo {title} {{A dynamical gluon mass solution in a coupled system of the Schwinger-Dyson equations}},}\ }\href {\doibase 10.1088/1126-6708/2004/08/057} {\bibfield  {journal} {\bibinfo  {journal} {JHEP}\ }\textbf {\bibinfo {volume} {08}},\ \bibinfo {pages} {057} (\bibinfo {year} {2004})},\ \Eprint {http://arxiv.org/abs/hep-ph/0408254} {arXiv:hep-ph/0408254} \BibitemShut {NoStop}%
%\%CITATION = HEP-PH/0408254;\%\%
\end{thebibliography}%

\onecolumngrid
\newpage
\begin{center}
  \textbf{\large \MakeUppercase{Supplemental Material}}
\end{center}

\setcounter{section}{0}
\section{Symmetries and anomalies}
\label{app:anomalies}
\begin{table}[h!]
\centering
\scalebox{1.3}{
\renewcommand{\arraystretch}{1.5}
\begin{tabular}{c||c||c|c} 
{\quad\quad} &\quad $SU(N_c)$\quad\quad & \quad$\, \,SU\left( N_c+4\right)\,\,$\quad \quad &\quad  $U(1)$ \quad  \\  
\hline
$\psi$ & $\overline{{\tiny\yng(1)}}$ & \tiny{$\yng(1)$}   &{ $-(N_c +2)$ } \\
$\chi$ & $\tiny\yng(2)$ & {$1$}   & {$ N_c+4$} \\

\end{tabular}
}
\caption{Fermion content and quantum numbers of the BY model. 
The $U(1)$ factor here is free from the $U(1)[SU(N_c)]^2$ Adler-Bell-Jackiw anomaly.}
\label{table:BYcontent}
\end{table}

In this Section, we review the symmetries of the BY theory, derive the anomaly-free global symmetry, and verify the anomaly matching through baryonic bound states. The theory contains two Weyl fermions, $\psi$ and $\chi$, transforming respectively in the anti-fundamental ($\tiny \overline{\yng(1)}$) and two-index symmetric (${\tiny\yng(2)}$) representations of $SU(N_c)$, see \Cref{table:BYcontent}. The two fermion species admit independent fermion-number symmetries $U(1)_\psi$ and $U(1)_\chi$. A generic linear combination defines a global $U(1)$ under which the fermions carry $q_\psi$ and $q_\chi$ charges. We keep these arbitrary and determine the anomaly-free combination by imposing the vanishing of the mixed gauge anomaly. Anomalies are encoded in the path-integral measure and can be computed from fermion triangle diagrams. They are characterised by group-theoretical coefficients given by fully symmetric traces over three generators (or charges for $U(1)$ symmetries), with each fermion species contributing according to its representation.\\

\noindent
\underline{\textbf{Gauge Anomalies:}} We first verify the cancellation of gauge anomalies. For the relevant $SU(N_c)$ representations, the cubic anomaly coefficients $A_R$ are defined from the trace of the symmetrised generators:
\begin{align}
   2\,\mathrm{Tr} \left[T_R^a\{T_R^bT_R^c\}\right]=A_R\ d^{abc}\,,
\end{align}
with $d^{abc}$ the ordinary totally symmetric invariant tensor, thus by definition $A_{\tiny\yng(1)}=1$. Specifically,  in the BY theory we have $A_{\tiny\overline{\yng(1)}}=-1$ and $A_{\tiny\yng(2)}=N_c+4$.
Hence, the $SU(N_c)^3$ gauge anomaly cancels:
\begin{align}
SU(N_c)^3:\qquad
(N_c+4)\,A_{\tiny\overline{\yng(1)}}+A_{\tiny\yng(2)}=0 \,.
\end{align}

We now consider the mixed gauged-$U(1)$ anomaly defined above, to which each fermion contributes with its representation index, charge and multiplicity. In the BY, it reads:
\begin{align}
SU(N_c)^2\times U(1):
\qquad
(N_c+4)\,T_{\tiny\overline{\yng(1)}}\,\,q_\psi + T_{\tiny{\yng(2)}}\,\,q_\chi
=
(N_c+4)\frac12 q_\psi+\frac{N_c+2}{2}q_\chi 
\end{align}

With $T_R$ the index of the $R$ representation. Requiring anomaly cancellation fixes the anomaly-free combination $ q_\chi=-\frac{N_c+4}{N_c+2}\,q_\psi$. The orthogonal combination remains anomalous, analogously to the axial $U(1)_A$ in QCD. For convenience we subsequently fix the overall normalization by choosing $ q_\psi=-(N_c+2)$, which yields the charge assignment shown in~\Cref{table:BYcontent}.\\

\noindent
\underline{\textbf{Flavour Anomalies:}} The resulting global symmetry is $SU(N_c+4)\times U(1)$, whose corresponding UV anomalies are:
\begin{subequations}
\begin{align}
SU(N_c+4)^3 &: \qquad N_c \\
SU(N_c+4)^2\times U(1) &: \qquad
N_c\,\frac12\,q_\psi \\
U(1)^3 &: \qquad
\frac{N_c(N_c+1)}{2}\,q_\chi^3 + N_c(N_c+4)\,q_\psi^3 =\frac{(N_c+4)(N_c+3)}{2}
\left(\frac{N_c}{N_c+2}q_\psi\right)^3
\end{align}
\end{subequations}

Gauge-singlet fermionic bound states can be constructed as $\mathcal B \sim \langle \chi\psi\psi\rangle$. Under the flavour group they decompose into symmetric and antisymmetric, $\mathcal B_{S/A}$, with IR anomalies:
\begin{subequations}
\begin{align}
SU(N_c+4)^3 &: \qquad
\mathcal B_S \rightarrow N_c+8
\qquad
\mathcal B_A \rightarrow N_c\\
SU(N_c+4)^2\times U(1) &: \qquad
\mathcal B_S \rightarrow \frac{N_c+6}{2}\,\frac{N_c}{N_c+2}q_\psi
\qquad
\mathcal B_A \rightarrow \frac{N_c+2}{2}\,\frac{N_c}{N_c+2}q_\psi\\
U(1)^3 &: \qquad
\mathcal B_S \rightarrow \frac{(N_c+4)(N_c+5)}{2}
\left(\frac{N_c}{N_c+2}q_\psi\right)^3
\qquad
\mathcal B_A \rightarrow \frac{(N_c+4)(N_c+3)}{2}
\left(\frac{N_c}{N_c+2}q_\psi\right)^3 
\end{align}
\end{subequations}

One finds that the antisymmetric baryon $\mathcal B_A$ reproduces the UV anomaly structure, providing an anomaly-matching solution in which confinement occurs without spontaneous symmetry breaking, as first derived in \cite{Bars:1981se}. 
\section{Effective action}
\label{app:effective action}
In this Section we provide further details on the effective action introduced in \labelcref{eq:fulleffectiveaction} and the approximation employed to it. The effective action and its scale dependent version includes all possible operators allowed the symmetries. This includes those present at the classical limit as well as generated by anomalies or by $\dSSB$. In the present work we employ the minimal truncation to the effective action known to qualitatively reproduce confining correlation functions and $\dSSB$. This consists of including up to all marginally relevant operators in the UV with their classical tensor structures together with four-fermion operators of canonical mass dimension six. Consequently, the approximation to the gauge sector reads,
\begin{align}\nonumber
		\Gamma_{\textrm{gauge},k}[A,c,\bar c]  =&\,
		\frac12 \int_p \, A^a_\mu(p) \,  \left[  \tilde Z_{A,k}\,\left( p^2+m^2_{\textrm{gap},k} \right)\Pi^\perp_{\mu\nu}(p) 
		+\frac{ 1}{\xi}Z^\parallel_{A,k}\,\left( p^2 + m^2_{\textrm{\tiny{mSTI}},k}\right)  \frac{p_\mu p_\nu}{p^2} \right] \, A^a_\nu(-p)  \\[2ex]\nonumber
		&\hspace{-2cm}  + \frac{1}{3!} \int_{p_1,p_2} \tilde Z_{A,k}^{3/2}\,\, g_{A^3,k}  \left[ {\cal T}_{A^3}^{(1)}(p_1,p_2)\right]^{a_1 a_2 a_3}_{\mu_1\mu_2\mu_3}
		\prod_{i=1}^3 A^{a_i}_{\mu_i}(p_i)  
		+  \frac{1}{4!}\int_{p_1,p_2,p_3} \hspace{-.3cm} \tilde Z_{A,k}^{2}\,\, g_{A^4,k}   \left[ {\cal T}_{A^4}^{(1)}(p_1,p_2,p_3)\right]^{a_1 a_2 a_3 a_4}_{\mu_1\mu_2\mu_3\mu_4}
		\prod_{i=1}^4 A^{a_i}_{\mu_i}(p_i)    \\[2ex]
		&\hspace{-2cm}	+ 	\int_p   Z_{c,k} \, \bar c^{\,a}(p)\,\, p^2 \delta^{ab}c^b(-p) + \int_{p_1,p_2} Z_{c,k} \tilde  Z_{A,k}^{1/2}\,\,g_{c \bar c A,k}  \left[ {\cal T}_{A\bar c c}^{(1)}(p_1,p_2)\right]^{a_1 a_2 a_3}_{\mu}
		\bar c^{a_2} (p_2) c^{a_1}(p_1) A^{a_3}_\mu(-p_1-p_2)   \,.
		\label{eq:gaugeeffaction}
\end{align}
Here, $\tilde Z_{A,k}$ and $Z^\parallel_{A,k}$ stand for the wave function renormalisation of the transversal and longitudinal gauge modes and  $Z_{c,k}$ for the ghost fields. $g_{i,k}$ are the different gauge coupling avatars and  $m^2_{\textrm{gap},k}= k^2 \bar m^2_{\textrm{gap},k}$ and $m^2_{\textrm{\tiny{mSTI}}}$ stand the gauge field mass gap in the transversal and longitudinal modes respectively. ${\cal T}_{\Phi_{i_1}\cdots \Phi_{i_n}}^{(1)}(p_1,...,p_n)$ label the standard classical tensor structures which can be found in previous works \cite{Ihssen:2024miv,Goertz:2024dnz,Cyrol:2016tym} and, accompanied by the vertex dressings $\tilde\Gamma^{(\Phi_{i_1}\cdots \Phi_{i_n})}(p_1,...,p_n)$, render the respective $n$-point functions,
\begin{align}
	\Gamma^{(\Phi_{i_1}\cdots \Phi_{i_n})}(p_1,...,p_n) =\frac{\delta}{\delta \Phi_{i_1}(p_1)}\cdots \frac{\delta}{\delta\Phi_{i_n}(p_n)}\Gamma[\Phi]=  \tilde\Gamma^{(\Phi_{i_1}\cdots \Phi_{i_n})}(p_1,...,p_n) \, \, {\cal T}^{(1)}_{\Phi_{i_1}\cdots \Phi_{i_n}}(p_1,...,p_n) \,.
	\label{eq:DefGn}
\end{align}
With these $n$-point function dressings we have defined the exchange couplings in \labelcref{eq:gaugeavatars,eq:gauge-fermionavatars}. Generally, these couplings are functions of the external momenta flowing through the vertex, but in the present work where these are computed at vanishing momenta we employ its dependence on the cutoff scale.

In this work we employ the Landau gauge, $\xi=0$, which is commonly used in computations of this type due to the simplifications it provides by removing longitudinal contributions. Moreover, it facilitates the control of deviations from the STIs, through the absence of longitudinal modes.
This therefore provides a setup that naturally improves the quality of the approximation. Functional studies have also been performed in other covariant gauges~\cite{Ferreira:2025tzo,Napetschnig:2021ria,Huber:2015ria,Aguilar:2015nqa,Alkofer:2003jr}. 

The gauge-fermion part of the effective action encodes the fermion dispersion relations as well as the gauge-fermion interactions,
\begin{align}
&\Gamma_{{\rm gauge-fermion},k } \left[A_\mu,\psi^\dagger, \psi,\chi^\dagger, \chi\right]=\\[1ex]
&\hspace{-0cm}\int_p  Z_{\psi,k}\,{\psi}^{\dagger\,, i,\, f}(p)\,\bar \sigma_\mu  {\partial}_\mu \,\psi^{i,\,f}(-p) - {\rm i}\int_{p_1,\,p_2} \tilde Z_{\psi,k}\tilde Z^{1/2}_{A,k}\,g_{A \psi^\dagger\psi,k } \, \,{\psi}^{\dagger\,, i_1 f_1}(p_2)\,\left[ {\cal T}_{A\psi^\dagger \psi}^{(1)}(p_1,p_2)\right]^{a \, i_1i_2f_1f_2}_{\mu }A^a_\mu(-p_1-p_2) \psi^{i_2f_2}(p_1)\notag\\[1ex]
&\hspace{-0cm}+\int_p Z_{\chi,k}\,{\chi}^{\dagger\,, \alpha}(p)\,\bar\sigma_\mu  {\partial}_\mu \chi^\alpha(-p)- {\rm i}\int_{p_1,\,p_2}Z_{\chi,k}\tilde Z^{1/2}_{A,k}\,g_{A \chi^\dagger\chi,k } \, {\chi}^{\dagger\,, \alpha_1}(p_2)\,\left[ {\cal T}_{A\chi^\dagger \chi}^{(1)}(p_1,p_2)\right]^{a \, \alpha_1\alpha_2}_{\mu } A^a_\mu(-p_1-p_2) \chi^{\alpha_2}(p_1)\notag
\label{eq:DiracAction}
\end{align}
where $\bar \sigma_\mu=(\mathbbm{1},-\boldsymbol{\sigma})$ with the Pauli matrices $\boldsymbol{\sigma}$ and the classical tensor structures
\begin{align}
    &\left[ {\cal T}_{A\psi^\dagger \psi}^{(1)}(p_1,p_2)\right]^{a\, i_1 i_2 f_1f_2}_{\mu }= \bar{\sigma}_\mu (T^{a}_{\overline{\rm {F}}})^{i_{1}i_{2}}\delta^{f_1f_2} &{\rm and}&&\left[ {\cal T}_{A\chi^\dagger \chi}^{(1)}(p_1,p_2)\right]^{a\, \alpha_1\alpha_2}_{\mu }= \bar{\sigma}_\mu (T^{a}_{\rm S})^{\alpha_{1}\alpha_{2}}_{}\,.
\end{align}
$T_{\overline{{\rm F}}}$ and $T^{a}_{\rm S}$ are the $SU(N)$ generators for anti-fundamental and two-indices symmetric representations respectively and ${f_i}$ stand for flavour indices.
$T_{\overline{{\rm F}}}$ is related to the generator in the fundamental representation $T^a_{\rm F}$ by,
\begin{align}
\left(T^a_{\overline{{\rm F}}}\right)^{i j}=-\left(T^a_{\rm F}\right)^{*i j}=-\left(T^a_{\rm F}\right)^{j i},
\end{align}
and in this work we use the convention ${\rm tr}(T^a_{\rm F}T^b_{\rm F})=\delta^{ab}/2$. 
In the above equation, the single-index notation $\chi^\alpha$ is used to denote a field transforming in the two-index symmetric representation. We deliberately employ a Greek index, $\alpha$, rather than a Roman index to indicate that it refers to the symmetric representation rather than the fundamental one. 
Using fundamental indices, $\chi$ can be written as $\chi^{\{ij\}}$, where the curly bracket represents the symmetrization of the two indices. 
One can make the identification between the two notation $\chi^\alpha\sim \chi^{\{ij\}}$, such that $\alpha$ takes the value from $1$ to $N(N+1)/2$ corresponding to the ordered tuples of two fundamental indices $(i,j)$ with $j\leq i$. In this way, we can formally express the $T^{a}_{\rm S}$ in terms of $T^a_{\rm F}$ as follows:
\begin{align}
    T^{a}_{\rm S}{}^{(i_1i_2)(j_1j_2)} =\frac{1}{2}\left((T^a_{\rm F})^{i_1i_2}\delta^{j_1j_2}+\delta^{i_1i_2}(T^a_{\rm F})^{j_1j_2}+(i_2\leftrightarrow j_2)\right), 
\end{align}
where $\delta$ is the Kronecker delta and the above equation serves as a basic input of our symbolic tracing of $SU(N)$ algebra.

In the purely fermionic sector of \labelcref{eq:four-fermioneff} we account for the lowest order (canonically most relevant) operators which are those with four fermions. For chiral gauge theories with two fermion species this was first derived in \cite{Li:2025tvu} and in this work the colour traces have been adapted for the BY theory:
\begin{subequations}\label{eq:4foperators}
\begin{align}
    {\cal O}^\psi_1& = ({\psi^\dagger}^{i_1f_1}\,\bar{\sigma}_\mu \,\psi_{i_1f_1 })\,({\psi^\dagger}^{i_2f_2}\,\bar{\sigma}_\mu \, \psi_{ i_2f_2})\,,\\[1.5ex]
    {\cal O}^\psi_2& = ({\psi^\dagger}^{i_1f_1}\,\bar{\sigma}_\mu \,\psi_{i_1f_2 })\,({\psi^\dagger}^{i_2f_2}\,\bar{\sigma}_\mu \, \psi_{ i_2f_1})\,,\\[1.5ex]
    {\cal O}^\chi_4 & =  ({\chi^\dagger}^{\alpha_1}\,\bar{\sigma}_\mu \, \chi_{  \alpha_1})\,({\chi^\dagger}^{\alpha_2}\,\bar{\sigma}_\mu \,\chi_{\alpha_2})\,,\\[1.5ex]
    {\cal O}^\chi_5 & =  ({\chi^\dagger}^{\alpha_1}\,\bar{\sigma}_\mu \, T_{\rm S}^{\alpha_1\alpha_2}\chi_{ \alpha_2})\,({\chi^\dagger}^{\alpha_3}\,\bar{\sigma}_\mu \,T_{\rm S}^{\alpha_3\alpha_4}\chi_{\alpha_4})\,,\\[1.5ex]
    {\cal O}^{\chi\psi}_6 & =  ({\psi^\dagger}^{i_1f_1}\,\bar{\sigma}_\mu \,\psi_{i_1 f_1})\,({\chi^\dagger}^{\alpha_2}\,\bar{\sigma}_\mu \, \chi_{ \alpha_2})\,,\\[1.5ex]
    {\cal O}^{\chi\psi}_7 & =  ({\psi^\dagger}^{i_1f_1}\,\bar{\sigma}_\mu \, T_{\rm \overline{F}}^{i_1i_2}\psi_{i_2 f_1 })\,({\chi^\dagger}^{\alpha_3}\,\bar{\sigma}_\mu \,T_{\rm S}^{\alpha_3\alpha_4}\chi_{ \alpha_4})\,.
\end{align}
\end{subequations}
We emphasise that $\chi$ and $\psi$ in the above operators are two-component left-handed Weyl spinors. We deliberately organise the Lorentz structure in terms of pairs of conjugated fields contracted with $\bar{\sigma}^\mu$. This makes explicit that each fermion bilinear is equivalent to $\overline{\chi}_L\gamma^\mu\chi_L$ in the four-component projected Dirac notation, while (pseudo-)scalar bilinears are forbidden since they require both left- and right-handed components.

When discussing the symmetry-breaking pattern and fermion condensates, it is convenient to rewrite the operators to avoid seemingly Lorentz-violating structures such as $\langle \chi^\dagger \chi \rangle$. Using the identity
$\bar{\sigma}^{\mu\dot{\alpha}\alpha}\,\bar{\sigma}_{\mu}^{\dot{\beta}\beta} = 2\,\epsilon^{\alpha\beta}\epsilon^{\dot{\alpha}\dot{\beta}}$
one can, for instance, express ${\cal O}^\chi_4$ as
\begin{align}
{\cal O}^\chi_4 = 2\,(\chi^{\dagger \alpha_1}\chi^{\dagger \alpha_2})(\chi_{\alpha_1}\chi_{\alpha_2})\,,
\end{align}
where each parenthesis denotes an $\epsilon$-contraction of two Weyl spinors and is manifestly Lorentz invariant. This differs from vector-like theories such as QCD, where the condensate is typically of the form $\langle \bar{\psi}\psi \rangle$.

For the sake of simplicity from now on we drop the $k$-argument of all dressings.
\section{RG-flows}
\label{app:RGflows}
Within the framework of the fRG, the progressive integration of momentum shells is implemented via the insertion of a sliding regulator at the level of the classical action. This enters bilinear in each of the fields and regularises their dispersion relation. For example, for the fields present in this work,
\begin{align}\label{eq:regulatorterm}
    \Delta S_k=\int_p A^a_\mu(p) R_{A,\,k}(p^2) \,A^a_\nu(-p) + \bar c(p) \,R_{c,\,k}(p^2) \,c(-p) + \psi^{\dagger,\,i\,,f}(p) R_{\psi,\,k}(p^2)\psi^{i\,,f}(-p)+\chi^{\dagger,\,\alpha}(p) R_{\chi,\,k}(p^2)\chi^\alpha(-p)\,.
\end{align}
Different regulator choices offer different advantages and can be used for optimisation. In the present work we employ the flat, or Litim \cite{Litim:2001up}, regulator, which allows for analytic integration over loop momenta and therefore leads to an analytic form of the flow equations. These read
\begin{subequations}
\begin{align}  
R_{A,\,k}(p^2) =& \,Z_{A}\left[ \Pi^\perp_{\mu\nu}(p) +\frac{1}{\xi}\frac{p_\mu p_\nu}{p^2}\right]\,  r_{\phi} (x),&	R_{c,\,k}(p^2) =& \,Z_{c}\, p^2 \,  r_{\phi} (x)&{\rm and}& &R_{\psi/\chi,\,k}(p^2) = \,{\rm i}\,  Z_{\psi/\chi}\,  \bar{\sigma}_\mu p_\mu \,  r_{\psi/\chi} (x),
	\label{eq:Regs}
\end{align}
for bosons and fermions respectively, with $x=p^2/k^2$, 
\begin{align}  
	r_{\phi} (x) =&  \,(1/x-1)\,\theta\left(1-x\right)&{\rm and}& &r_{\psi/\chi} (x) = \, (1/\sqrt{x}-1)\,\theta\left(1-x\right)\,.
	\label{eq:regxdeff}
\end{align}
\end{subequations}
As its apparent in \labelcref{eq:regulatorterm}, as the fermion propagator is proportional to its chiral dispersion, the regularisation scheme employed here does not pose any inconvenience in the treatment of chiral fermions.

The RG flows derived in this work are provided in an ancillary \texttt{Mathematica} notebook. They correspond to the set of couplings
\begin{align}\notag
 \{g_{A^3},g_{A^4},g_{A \bar c c },g_{A \chi^\dagger \chi},g_{A \psi ^\dagger \psi}, \bar m_{\rm gap}, \bar \lambda_1,\bar \lambda_2,\bar \lambda_4,\bar \lambda_5,\bar \lambda_6,\bar \lambda_7,\,\tilde Z_A,Z_c,Z_\psi,Z_\chi\}.   
\end{align}
These are derived from the  flow of the corresponding $n$-point functions, projected onto the classical tensor structures and evaluated in the symmetric momentum configuration at vanishing external momenta. For further details on the derivation of the flow equations we refer to the explicit computations in~\cite{Goertz:2024dnz,Li:2025tvu}.

\section{Details on confinement and the bootstrap approach}\label{app:easyconfinement}
As previously mentioned, in the present work we employ the minimal (and computationally simplest) truncation to the effective average action so far known to reproduce reliable confining correlation functions~\cite{Goertz:2024dnz}. This consists of the gauge sector introduced in \labelcref{eq:gaugeeffaction} together with a convenient parametrisation of the gauge field wave function,
\begin{align}
 Z_{A}=  \tilde Z_{A} (1+ \bar m_{{\rm gap}}^2)\,.
\end{align}
This approximation incorporates the power-law scaling generated with the emergence of the gauge mass gap and significantly reduces the computational cost of evaluating confining correlation functions. Nonetheless, as the STIs are recovered at the $k\to0$ limit the present treatment requires special care when reading off correlation functions at non-vanishing cutoffs. In this section we elaborate on this point.

The longitudinal gauge-field mass, $m^2_{\textrm{\tiny mSTI},k}$, is induced by the mSTIs in the presence of a momentum cutoff. In the physical limit $k\to 0$, the mSTIs reduce to the standard STIs in Landau gauge, and the longitudinal mass vanishes, $m^2_{\textrm{\tiny mSTI},k=0}=0$.
Hence, $m^2_{\textrm{\tiny mSTI},k}$ does not indicate a massive Yang--Mills theory, but rather a  modification of the Landau-gauge STIs due to the IR regulator~\cite{Cyrol:2016tym,Dupuis:2020fhh}. On the other hand, the transverse gauge-field mass gap remains non-zero, $m^2_{\textrm{gap},k\to0}\neq 0\,,$ which signals confinement. 

For cutoff scales above the confinement scale, where no physical mass is generated, the two masses coincide,$m^2_{\textrm{gap},k} = m^2_{\textrm{\tiny mSTI},k}\,,$ since their difference can only arise from irregular vertices~\cite{Cyrol:2016tym}. By confinement scale we refer to the cutoff scale at which confining dynamics set in and longitudinal vertices develop massless modes through the underlying mechanism for confinement, eg. the Schwinger~\cite{Aguilar:2011xe,Aguilar:2021uwa,Ferreira:2023fva,Aguilar:2022thg} or  BRST quartet~\cite{Alkofer:2011pe}  mechanisms. In this regime, the flow of the transverse mass gap is given by
\begin{subequations}
\begin{align}
k \partial_k \left( \tilde Z_{A}\, m^2_{{\rm gap}} \right) 
&= 
\left. \frac{{\rm tr}\!\left[ \Pi^\perp_{\mu\nu} \delta^{ab} \, k \partial_k\,\Gamma^{(AA)}_k\right]}{3(N_c^2-1)}\right|_{p^2=0}
=
\tilde Z_A k^2\, \overline{\textrm{Flow}}_{AA}(0)\,\qquad\qquad \qquad\qquad\qquad{\rm with,}
\\
\overline{\textrm{Flow}}_{AA}(0)
&= \frac{1}{4(4\pi)^2}\Bigg[
-\frac{8 N_c \, g_{A\bar c c }^2}{6}\!\left(1-\frac{\eta_c}{8} \right)
+\frac{8 N_c \, g_{A^3 }^2}{\left(1+\bar m_{\rm gap}^2\right)^3}\!\left(1-\frac{\eta_A}{8} \right)
-\frac{18 N_c \,  g_{A^4 }^2}{2 \left(1+\bar m_{\rm gap}^2\right)^2}\!\left(1-\frac{\eta_A}{6} \right)
\\
&\hspace{5.5cm}\notag
+\,2 (N_c+2) \, g_{A  \psi^\dagger \psi}^2\!\left(1-\frac{\eta_\psi}{5} \right)
+\,2 (N_c+4) \, g_{A \chi^\dagger \chi}^2\!\left(1-\frac{\eta_\chi}{5} \right)
\Bigg] .
\end{align}
\end{subequations}
Here we have separated the dimensionful factor $\tilde Z_A k^2$, where $\tilde Z_A$ carries the logarithmic RG running of the gauge two-point function. The remaining contribution, $\overline{\textrm{Flow}}_{AA}(0)$ is evaluated at vanishing momentum and carries the power-law scaling corrections, which accounts for the generation of the physical mass gap as well as of $m_{\rm mSTI}^2$.

In full momentum-dependent computations, the deviation from the STIs is minimised since the flow is integrated down to $k\to 0$, and external momenta select the relevant contributions. For instance, tadpole diagrams with four-gauge interactions are present at the cutoff scale but vanish after renormalisation and due to momentum independence.

In the present truncation without momentum dependencies, the deviation from the STIs must be controlled explicitly. If they are present at scales higher than the confining one, such contributions would merely reflect the presence of the regulator.
However, on the other hand, it is important to stress that the power-law corrections are required in the IR: they are compatible with the STIs and encode the mechanism underlying confinement. In the present approach the control is achieved by including the power-law flow only in the regime where the physical mechanism generating the mass gap is active. 
The onset scale for the power-law corrections is chosen close to the confinement scale, signalled by the peak of the gauge propagator. Its important to note that the appearance of a gapping in the wave function does not qualitatively depend on the onset of the power-law corrections.

\begin{figure}
    \centering
    \includegraphics[width=.3\columnwidth]{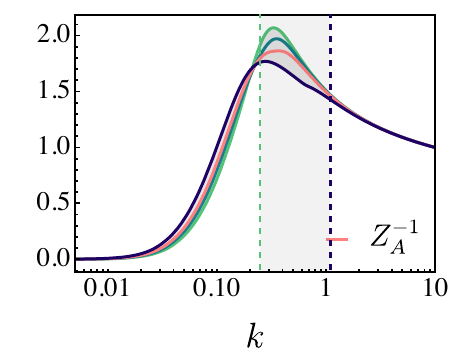}
    \includegraphics[width=.3\columnwidth]{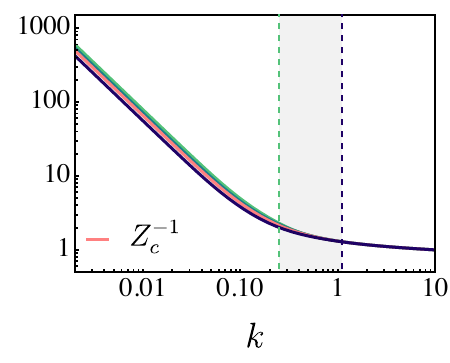}
    \includegraphics[width=.3\columnwidth]{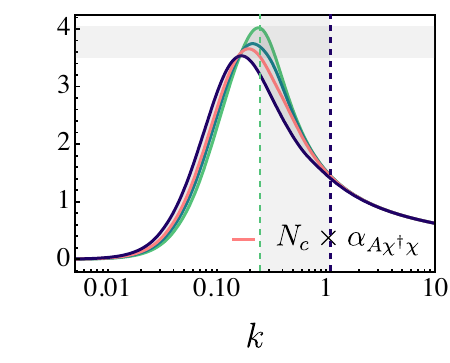}
    \caption{Gauge (left) and ghost (centre) field wave functions and the $\chi$-gauge exchange coupling (right) in a $N_c=5$ BY theory with different onsets of the power-law scaling in the flow of the mass gap. From darker blue to green the onset (given by the deviation from the logarithmic scaling) is delayed between to limiting scenarios as marked by the gray vertical shaded area. The horizontal shaded area in the right-most plot shows the error estimate in the peak of the gauge-fermion strength.
    }
    \label{fig:OnsetPF}
\end{figure}

In \Cref{fig:OnsetPF}, we show integrated RG flows for a BY theory with $N_c=5$ and different choices for the onset scale of the power-law correction. We vary this scale between two extreme cases: an onset at high scales, where power-law terms affect the perturbative RG scaling (vertical blue dashed line), and an onset below the confinement scale, i.e. after the peak of the gauge propagator (vertical green dashed line). For the main results of this work, we employ onset scales that preserve the perturbative running while incorporating power-law scaling at the confinement scale. In the right panel of \Cref{fig:OnsetPF}, we display the gauge--$\chi$ exchange couplings; the variation of their peak values is included in the error estimate quoted in the main text and shown in \Cref{fig:alphacrit}. Further details of the present approach can be found in \cite{Goertz:2024dnz}.

\begin{figure}
    \centering
    \includegraphics[width=.45\columnwidth]{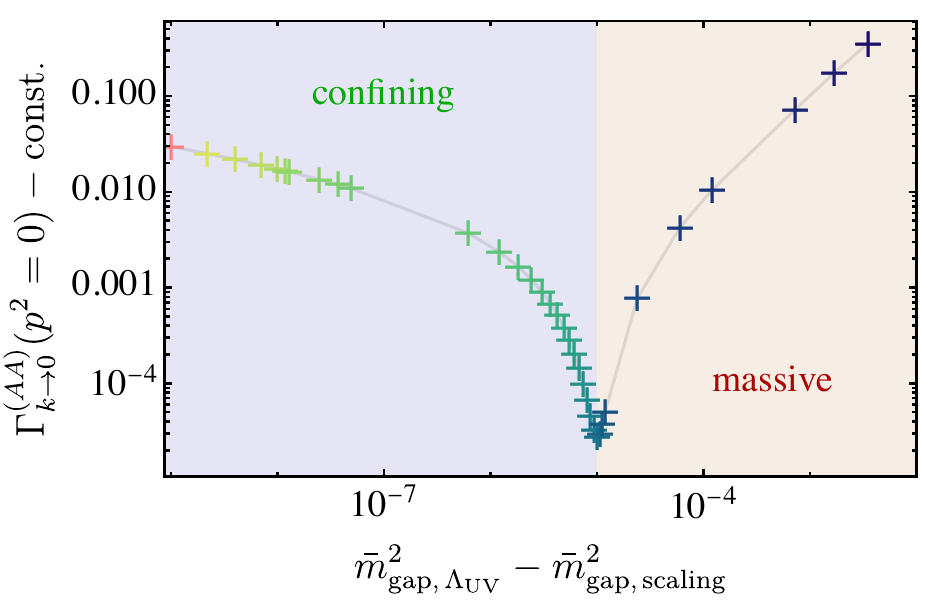}\\[2ex]
    \includegraphics[width=.32\columnwidth]{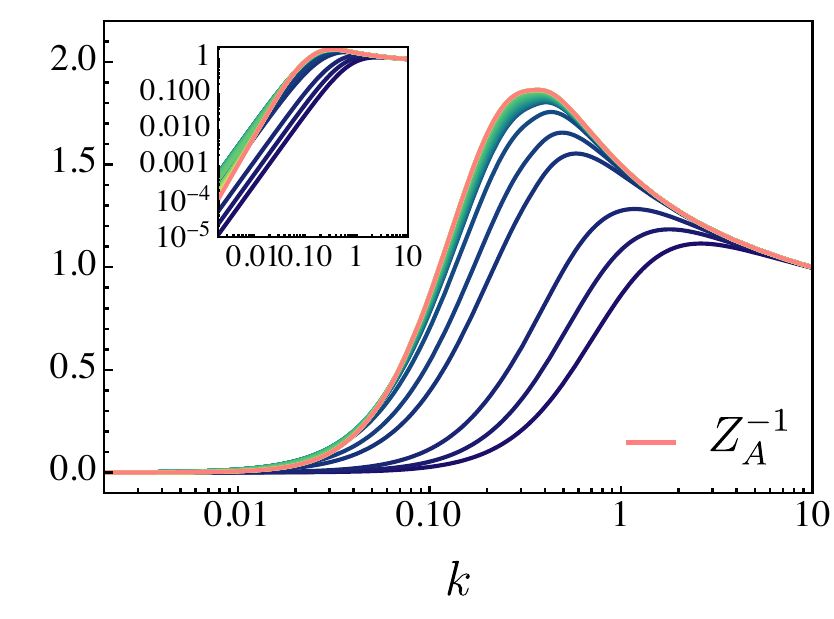}
    \includegraphics[width=.32\columnwidth]{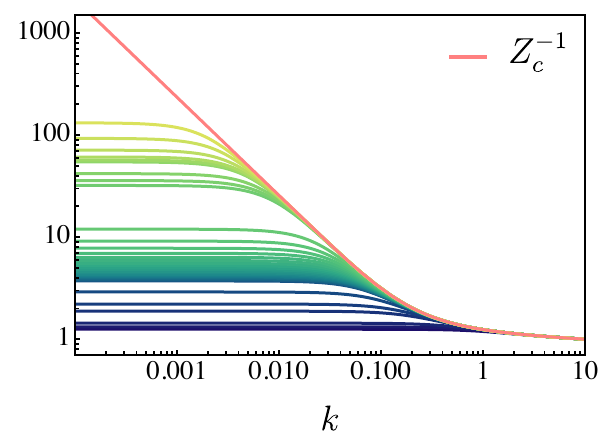}
    \includegraphics[width=.32\columnwidth]{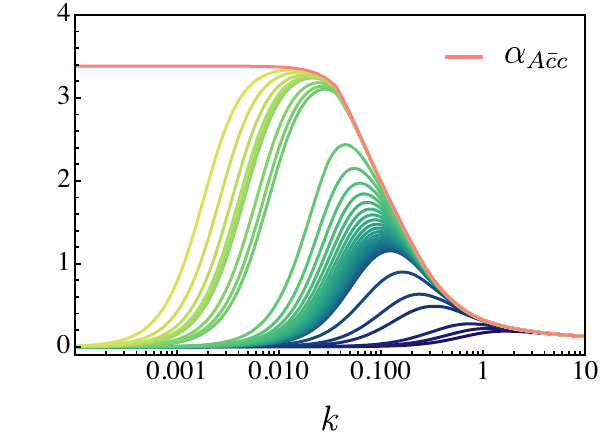}
    \caption{ On the top, we show the value of the gauge field two point function in the deep IR as a function of the value of the mass gap in the UV. 
    On the bottom row, we show the gauge (left) and ghost (centre) wave function and the ghost-gauge exchange coupling (right) for a $N_c=5$ BY theory. Different curves correspond to different onsets of the quadratic part of the gauge mass gap with the same colour coding as the top plot. From darker blue to green the onset (given by the deviation from the logarithmic scaling) is delayed. }
    \label{fig:tunningconf}
\end{figure}

Beyond the specific approximation used in this work, the bootstrap implementation in functional approaches allows to determine confining correlation functions. The Kugo-Ojima criterion assumes the existence of a global BRST transformation and colourless asymptotic states in the physical Hilbert space. In the Landau gauge, this criterion translates into a unique scaling of correlation functions in the deep IR, with $Z_A \propto (p^2)^{-2\kappa}$, $Z_c \propto (p^2)^{\kappa}$ and $1/2<\kappa<1$ \cite{Cyrol:2016tym,Kugo:1995km,Alkofer:2000wg}. This is equivalent to a gauge-suppressed propagator given the presence of a scaling mass gap and a IR-enhanced ghost propagator. In the fRG the Kugo-Ojima criterion translates into a unique value of the $\bar m_{\rm gap}^2$ at the UV boundary. In \Cref{fig:tunningconf} we depict  RG-trajectories for different boundary values of the mass gap. The pink line shows the IR scaling solution which is employed in the results shown in this work. Furthermore, different values of this mass parameter allow tuning different solutions such as the decoupling solution \cite{Cyrol:2016tym,Aguilar:2008xm,Aguilar:2004sw} and massive Yang-Mills solutions. 

\section{ Dynamical symmetry breaking and dominant four-fermion channel}
In this appendix we provide further details on the fermionic dynamics responsible for the onset of $\dSSB$ in the Bars--Yankielowicz class of chiral gauge theories. In particular, we identify the dominant four-fermion channel that controls the fixed-point merger underlying the divergence of the RG flow.

The mechanism leading to $\dSSB$ in the fRG framework is governed by the fixed-point merger in the flow of four-fermion couplings. As the gauge interaction grows along the RG flow, the interacting fixed point of the four-fermion system approaches the near-Gaussian one and can eventually merge with it. Beyond this critical strength, the RG flow no longer admits real fixed-point solutions and the four-fermion coupling develops runaway behaviour.

The minimal gauge strength required to induce this fixed-point merger defines the critical coupling $\acrit$. The general structure of this mechanism and its relation to $\dSSB$ have been discussed in detail in our previous analysis of Georgi--Glashow theories~\cite{Li:2025tvu}. We refer the interested reader to that work for a more comprehensive discussion. As mentioned above, the four-fermion flows are organised in the attached \texttt{Mathematica} notebook.

In the BY class we determine $\acrit$ by gradually increasing $\alpha_g$ from zero and solving the coupled system of RG flow equations for the four-fermion sector until a singularity appears at finite RG time, signalling the fixed-point merger. We find that the critical value is already reproduced when restricting the system to the two pure $\chi$ operators associated with the couplings $\lambda_4$ and $\lambda_5$. This observation shows that the instability leading to $\dSSB$ originates entirely in the $\chi$ sector and that the other channels do not quantitatively contribute. Consequently, the condensate expected to form is 
$\langle \chi\chi\rangle$, whereas identifying the detailed symmetry-breaking pattern and the direction of the condensate within the multiplet requires a bosonized description of the theory. The operators ${\cal O}^\chi_4$ and ${\cal O}^\chi_5$ therefore span the relevant subspace in which the dominant fermionic dynamics takes place.

A closer inspection of the flows of  ${\cal O}^\chi_4$ and ${\cal O}^\chi_5$ shows that only ${\cal O}^\chi_5$ exhibits the fixed-point merger. In addition, for $N_c\geq 3$, this merger crucially depends on the evolution of $\lambda_4$ through four-fermion diagrams with insertions of ${\cal O}^\chi_4$; without these contributions $\acrit$ changes (increases) significantly. This motivates the construction of a basis in which the RG flow is dominated by a single effective operator. Here we therefore rotate the operator basis by diagonalising the mixing coefficients $\boldsymbol{c}^{B}$ in the subsector spanned by ${\cal O}^\chi_4$ and ${\cal O}^\chi_5$.

In the original basis, $\boldsymbol{c}^{\rm B}_{ij}$ with $i,j=4,5$ reads
\begin{align}
  \boldsymbol{c}^{\rm B}_{ij}=  \frac{-5 \eta _A-6 \eta _{\chi }+60}{20 \pi }\left(
\begin{array}{cc}
 0 & 2 \left(N_c^2+N_c-2\right)/N_c^2 \\
 1 & 1-4/N_c \\
\end{array}
\right)\,.
\end{align}
After diagonalisation the operators mix, leading to a rotated basis $\tilde{\cal O}^\chi_4$ and $\tilde{\cal O}^\chi_5$ defined by the rotation matrix
\begin{align}
  R =  \left({N_c \left(5 N_c-8\right)+4}\right)^{-1/2}\left(
\begin{array}{cc}
 {2-2 N_c} & {N_c} \\
 {N_c+2} & {N_c} \\
\end{array}
\right),
\end{align}
such that
\begin{align}
   & \tilde{\boldsymbol{c}}^{\rm B}_{ij}=R\,\boldsymbol{c}^{\rm B}_{ij}\,R^{-1}
   ={\rm diag}\!\left[\frac{2}{N_c}-2,\frac{2}{N_c}+1\right]\,\,
   \frac{-5 \eta _A-6 \eta _{\chi }+60}{20 \pi } &{\rm and ,}&&
   &\lambda_i= R_{ij}\tilde{\lambda}_j,\qquad 
    {\cal O}_i = (R^{-1})^T \tilde{\cal O}_j .
\end{align}

With this diagonalisation the coefficients $\tilde{\boldsymbol{c}}^{\rm C}_{lij}$ with $l,i,j=4,5$ become
\begin{align}
  \tilde{\boldsymbol{c}}^{\rm C}_{4ij}&=  \frac{5-\eta_\chi}{40\pi^2 N_c}\left(
\begin{array}{cc}
  N_c\left(-N_c^2-N_c+2\right)/2 & -{N_c^2+N_c-2}\\
 -N_c^2+N_c-2 & 6 \left(N_c^2+N_c-2\right)/N_c \\
\end{array}
\right)&{\rm and, }&&
 \tilde{\boldsymbol{c}}^{\rm C}_{5ij}&=  \frac{5-\eta_\chi}{40\pi^2}\left(
\begin{array}{cc}
 0 & 2 \\
 2 & -2 \left(N_c^2+4\right)/N_c \\
\end{array}
\right).
\end{align}

In this rotated basis we find that the operator
\begin{align}
\label{eq:Odom}
   \tilde {\cal O}^\chi_{5}= 
   \frac{(2-2 N_c)\,{\cal O}^\chi_4 +N_c\,{\cal O}^\chi_5}
   {\sqrt{N_c \left(5 N_c-8\right)+4}}
\end{align}
is the dominant channel driving the fixed-point merger, with only a minor influence from $\tilde{\cal O}^\chi_4$. 

We conclude by summarising that this non-unique diagonalisation provides a better understanding of the relevant channel giving rise to the highest symmetry breaking condensate.
\end{document}